\documentclass[conference,twocolumn]{IEEEtran}

\PassOptionsToPackage{bookmarks=false}{hyperref}

\usepackage{acronym}
\usepackage{acronym.conferences.telecom}
\usepackage{graphicx}

\usepackage{amsmath,amssymb}
\usepackage{enumerate}
\usepackage{pgfplots}
\usepackage{standalone}
\usepackage{tikz}
\usetikzlibrary{arrows,
	chains,
	decorations.markings,
	shadows, shapes.arrows,
	automata,positioning,
}

\newacronym{BER}{Bit Error Rate}
\newacronym{BS}{Base Station}
\newacronym{CIR}{Channel Impulse Response}
\newacronym{CFR}{Channel Frequency Response}
\newacronym{CP}{Cyclic-Prefix}
\newacronym{DF}{Decision-Feedback}
\newacronym{FDE}{Frequency Domain Equalization}
\newacronym{FFT}{Fast Fourier Transform}
\newacronym{IDFT}{Inverse Discrete Fourier Transform}
\newacronym{IBER}{Irreducible BER}
\newacronym{ISI}{Inter-Symbol Interference}
\newacronym{LOS}{Line-Of-Sight}
\newacronym{LDB}{Linear Detection Bound}
\newacronym{MF}{Matched Filter}
\newacronym{MFB}{Matched-Filter Bound}
\newacronym{ML}{Maximum Likelihood}
\newacronym{MLD}{Maximum-Likelihood Detection}
\newacronym{MLDB}{Maximum-Likelihood Detection Bound}
\newacronym{MIMO}{Multi-Input Multi-Output}
\newacronym{MRC}{Maximal Ratio Combining}
\newacronym{MU}{Multi-User}
\newacronym{MUI}{Multi-User Interference}
\newacronym{MSI}{Multi-Stream Interference}
\newacronym{MUD}{MultiUser Detection}
\newacronym{MT}{Mobile Terminals}
\newacronym{MMSE}{Minimum Mean-Squared Error}
\newacronym{OFDM}{Orthogonal Frequency Division Multiplexing}
\newacronym{QAM}{Quadrature Amplitude Modulation}
\newacronym{SINR}{Signal-to-Interference-plus-Noise-Ratio}
\newacronym{SC}{Single Carrier}
\newacronym{SU}{Single User}
\newacronym{TX}{Transmitter}
\newacronym{RX}{Receiver}
\newacronym{SIMO}{Single-Input Multi-Output}
\newacronym{SISO}{Single-Input Single-Output}
\newacronym{ZF}{Zero Forcing}

\begin{document}
\title{Detection Performance with Many Antennas Available for Bandwidth-Efficient Uplink Transmission in MU-MIMO Systems}

\author{$^{(1)}$Paulo Torres, $^{(2)}$Antonio Gusmao\\
$^{(1)}$IPCB - Instituto Politecnico de Castelo Branco, Portugal \\
$^{(2)}$IST - Instituto Superior Tecnico, Universidade de Lisboa, Portugal \\
$^{(1)}$paulo.torres@ipcb.pt, $^{(2)}$gus@ist.utl.pt}

\maketitle

\begin{abstract}

This paper is concerned with \SC/\FDE\ for bandwidth-efficient uplink block transmission, with \QAM\ schemes, in a \MU\ \MIMO\ system. The number of \BS\ receiver antennas is assumed to be large, but not necessarily much larger than the overall number of transmitter antennas jointly using the same time/frequency resource at \MT.

In this context, we consider several detection techniques and evaluate, in detail, the corresponding detection performances (discussed with the help of selected performance bounds), for a range of values regarding the number of available \BS\ receiver antennas. From our performance results, we conclude that simple linear detection techniques, designed to avoid the need of complex matrix inversions, can lead to unacceptably high error floor levels. However, by combining the use of such simple linear detectors with an appropriate interference cancellation procedure - within an iterative \DF\ technique -, a close approximation to the \SIMO\ \MFB\ performance can be achieved after a few iterations, even for 64-\QAM\ schemes, when the number of \BS\ antennas is, at least, five times higher than the number of antennas which are jointly used at the user terminals.

\end{abstract}


\section{Introduction}\label{sec_introd}

\CP-assisted  block transmission schemes, proposed and developed for broadband wireless systems, take advantage of current low-cost, flexible, \FFT-based signal processing technology, with both \OFDM\ and \SC/\FDE\ \cite{GC00}. Mixed air interface solutions (\OFDM\ for downlink and \SC/\FDE\ for uplink \cite{GC00}) are now widely accepted.

The development of \MIMO\ technologies has been crucial for the "success story" of broadband wireless communications. Through spatial multiplexing, following and extending ideas early presented in \cite{Foschini03}, \MIMO\ systems are currently able to provide very high bandwidth efficiencies and a reliable radio transmission at data rates beyond $1$ Gigabit/s. In the last decade, \MU-\MIMO\ systems have been successfully implemented and introduced in several broadband communication standards \cite{Gesbert07}; in such "space division mutiple access" systems, the more antennas the BS is equipped with, the more users can jointly communicate in the same time-frequency resource.

The adoption of \MU-\MIMO\ systems with a very large number of antennas in the BS, much larger than the number of \MT\ antennas in its cell, was proposed in \cite{Marzetta10}. This "massive \MIMO" approach was shown to be recommendable \cite{Marzetta10,Rusek13,Hoydis13}: simple linear processing for \MIMO\ detection/precoding (uplink/downlink), becomes quasi-optimal; both \MUI/\MSI\ effects and fast fading effects of multipath propagation tend to disappear; power and bandwidth efficiencies are substantially increased.

This paper deals with \SC/\FDE\ for the uplink of a \MU-\MIMO\ system where a high bandwidth efficiency is intended, through \QAM\ transmission schemes (up to 6 bits/symbol), and the \BS\ is constrained to adopt low-complexity detection techniques, but can be equipped with a large number of receiver antennas, not necessarily much larger than the overall number of transmitter antennas jointly using the same time/frequency resource at mobile terminals. In this context, either a linear detection or a reduced-complexity iterative \DF\ detection are considered, as presented in sec. \ref{sec2}: as to the linear detection alternative, we include the optimum, \MMSE\ detection \cite{Kim2008}, a reduced-complexity \MMSE-type detection \cite{ISCAS2013_wu} and the quite simple \MF\ detection; the iterative DF detection alternative, which resorts to joint cancellation of estimated \MUI/\MSI, is an extension of the iterative \DF\ technique considered by the authors, for 4-\QAM\ transmission, in \cite{PT2015VTC}, and can also be regarded as an extension to the multi-input context of reduced-complexity iterative receiver techniques proposed by the authors for \SIMO\ systems (see \cite{GC2007_Revista2} and the references therein).

This paper is an extended version of \cite{PT16_1}. We evaluate, in detail, the detection performances for the several detection alternatives (discussed with the help of selected performance bounds, as presented in sec. \ref{sec3}), for a range of values regarding the number of available BS receiver antennas. A wide set of numerical performance results and the main conclusions of the paper are presented in sections \ref{sec4} and \ref{sec5}, respectively.

\section{MU-MIMO Uplink System Model and Detection Techniques}\label{sec2}

\subsection{\SC/\FDE\ with \QAM\ Schemes for Block Transmission}\label{subsec2A}

We consider here a \CP-assisted \SC/\FDE\ block transmission, within a \MU-\MIMO\ system with $N_T$ TX antennas and $N_R$ RX antennas; for example, but not necessarily, one antenna per \MT. We assume, in the $j$th TX antenna ($j=1,2,...,N_T$) a length-$N$ block $\bold{s}^{(j)}=[s_0^{(j)}, s_1^{(j)},...,s_{N-1}^{(j)}]^T$  of time-domain data symbols in accordance with the corresponding binary data block and the selected QAM constellation, concerning $2m(j) $  bit/symbol, under a Gray mapping rule. A length-$L_s$ \CP, long enough to cope with the time-dispersive effects of multipath propagation, is also assumed.

By using the frequency-domain version of the time-domain data block $\bold{s}^{(j)}$, given by $\bold{S}^{(j)}=\left[S_0^{(j)},S_1^{(j)},\cdots,S_{N-1}^{(j)}\right]^T = DFT\left(\bold{s}^{(j)}\right)$ $(j=1,2,\cdots,N_T)$, we can describe the frequency-domain transmission rule as follows, for any subchannel $k$ $\left(k=0,1,\cdots, N-1\right)$:
\begin{eqnarray}\label{eq1}
   \bold{Y}_k = \bold{H}_k \bold{S}_k + \bold{N}_k,
\end{eqnarray}
where $\bold{S}_k = \left[S_k^{(1)},S_k^{(2)},\cdots,S_k^{(N_T)}\right]^T$ is the ''input vector'', $\bold{N}_k = \left[N_k^{(1)},N_k^{(2)},\cdots,N_k^{(N_R)}\right]^T$ is the Gaussian noise vector $\left(E\left[N_k^{(i)}\right]=0\right.$ and $\left.E\left[\left|N_k^{(i)}\right|^2\right]=\sigma_N^2=N_0 N\right)$, $\bold{H}_k$ denotes the $N_R\times N_T$ channel matrix with entries $H_k^{(i,j)}$, concerning a given channel realization, and $\bold{Y}_k= \left[Y_k^{(1)},Y_k^{(2)},\cdots,Y_k^{(N_R)}\right]^T$ is the resulting, frequency-domain, ''output vector'' .

As to a given \MIMO\ channel realization, it should be noted that the \CFR\ $\bold{H}^{(i,j)}=\left[H_0^{(i,j)}, H_1^{(i,j)},...,H_{N-1}^{(i,j)}\right]^T$, concerning the antenna pair $(i,j)$, is the DFT of the \CIR\ $\bold{h}^{(i,j)}=\left[h_0^{(i,j)}, h_1^{(i,j)},...,h_{N-1}^{(i,j)}\right]^T$, where $h_n^{(i,j)}=0$ for $n > Ls\ (n=0,1,...,N-1)$. Regarding a statistical channel model - which encompasses all possible channel realizations -, let us assume that $E\left[h_{n}^{(i,j)}\right]=0$ and $E\left[h_n^{(i,j)*}h_{n'}^{(i,j)}\right]=0$ for $n'\neq n$. By also assuming, for any $(i,j,k)$, a constant
\begin{eqnarray}\label{eq2}
   E\left[\left|H_k^{(i,j)}\right|^2\right] = \sum_{n=0}^{N-1} E\left[ \left| h_n^{(i,j)} \right|^2 \right] = P_\Sigma
\end{eqnarray}
(of course, with $h_n^{(i,j)}=0$ for $n>L_s$), the average bit energy, regarding the $jth$ \TX\ antenna, at each \BS\ antenna, is given by
\begin{eqnarray}\label{eq3}
   E_b^{(j)} = \frac{\left[\sigma_s^{(j)}\right]^2}{2\eta m(j)} P_\Sigma,
\end{eqnarray}
where $\eta=\frac{N}{N+L_s}$  and $\left[\sigma_s^{(j)}\right]^2=E\left[\left|s_n^{(j)}\right|^2\right]$.

\subsection{Linear Detection Techniques}\label{subsec2B}

\begin{figure}[!ht]
 	\centering
 		
\tikzstyle{block} = [draw, fill=white, rectangle, minimum height=1.5cm, minimum width=3cm,]	
\resizebox {\columnwidth} {!} {
 	\begin{tikzpicture}[
auto, node distance=2.35cm,
start chain = A going right,
every join/.style = {draw, -stealth, thick},
block/.append style = {on chain=A},
font=\huge,
]

\node (a) at (0,0) {};
\node [label={[xshift=.75cm, yshift=0.4cm] $\textbf{Y}_{k}$}] {};
\node [single arrow,minimum size=1.25cm,draw=black, fill opacity=1,minimum height=5em, outer sep=0pt,line width=0.0625cm,right=0pt of a.east] {};
\node [block,line width=0.125cm, fill opacity=1,right of=input,node distance=.45cm,font=\Huge]  (b) at (3,0) {$\widehat{\textbf{H}}^H_{k}$};
\node [single arrow,minimum size=1.25cm,draw=black, fill opacity=1,minimum height=4cm,outer sep=0pt,line width=0.0625cm,right=0pt of A-1.east] {};
\node [label={[xshift=6.75cm, yshift=0.5cm] $\widetilde{\textbf{Y}}_{k,MF}=\widehat{\textbf{H}}^H_{k}\textbf{Y}_{k}$}] {};
\node [block, line width=0.125cm, fill opacity=1,right of=input,node distance=6.2cm,font=\Huge] (c) at (2,0) {$\textbf{B}_k$};
\node [single arrow,minimum size=1.25cm,draw=black, fill opacity=1,minimum height=3.5cm, outer sep=0pt,line width=0.0625cm,right=0pt of A-2.east] {};
\node [label={[xshift=14.0cm, yshift=0.5cm] $\widetilde{\textbf{Y}}_{k} = \textbf{B}_k\widetilde{\textbf{Y}}_{k,MF}$}] {};
\node [font=\huge] (d) at (3.5,-1.25) {\bf MF/MRC};
\node[state,draw=none] (a1)  at (.8,.9) {};
\node[state,draw=none] (a2)  at (6.5,.9) {};
\node[state,draw=none] (a3)  at (13.4,.9) {};
\node[state,draw=none] (a4)  at (0,-1.2) {};
\begin{pgfinterruptboundingbox}
\node[state,draw=none] (b1) [below of=a1] {$\bf{N_R}$};
\node[state,draw=none] (b2) [below of=a2] {$\bf{N_T}$};
\node[state,draw=none] (b3) [below of=a3] {$\bf{N_T}$};
\end{pgfinterruptboundingbox}
\draw [-,dashed,line width=0.1cm] (a1) to[below]  node[auto] {} (b1);
\draw [-,dashed,line width=0.1cm] (a2) to[below]  node[auto] {} (b2);
\draw [-,dashed,line width=0.1cm] (a3) to[below]  node[auto] {} (b3);

\end{tikzpicture}
}\vspace{-.25cm}
 	\caption{Frequency-domain linear detection.}   \label{OFDM_MIMO_Block_1}
 \end{figure}
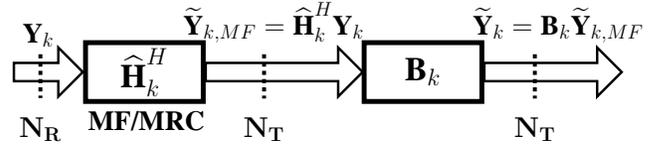

An appropriate linear detector can be implemented by resorting to frequency-domain processing. After CP removal, a DFT operation leads to the required set $\left\{\bold{Y}_k; \ k=0,1,\cdots,N-1\right\}$ of length-$N_R$ inputs to the frequency-domain detector ($\bold{Y}_k$ given by (\ref{eq1})); it works, for each $k$, as shown in Fig. \ref{OFDM_MIMO_Block_1}, leading to a set $\left\{\bold{\widetilde{Y}}_k; \ k=0,1,\cdots,N-1\right\}$ of length-$N_T$ outputs $\bold{\tilde{Y}}_k=\left[\tilde{Y}_k^{(1)},\tilde{Y}_k^{(2)},\cdots,\tilde{Y}_k^{(N_T)}\right]^T \ (k=0,1,\cdots,N-1)$.

When $N_T \leq N_R$, possibly with $N_R \gg 1$, either an \MMSE, frequency-domain, optimum linear detection or a reduced-complexity,  frequency-domain, linear detection can be considered. In all cases, the detection matrix, for each subchannel $k$ ($k=0,1,...,N-1$) can be written as
\begin{eqnarray}\label{eq4}
\textbf{D}_k =  \textbf{B}_k \widehat{\textbf{H}}_k^H,
\end{eqnarray}
where $\widehat{\textbf{H}}_k^H$  is the conjugate transpose of the estimated \MU-\MIMO\ channel matrix $\widehat{\bold{H}}_k$ and $\bold{B}_k$ is a selected $N_T\times N_T$ matrix, possibly depending on $\widehat{\textbf{H}}_k$.  Therefore, $\bold{\widetilde{Y}}_k = \bold{D}_k \bold{Y}_k = \bold{B}_k \widehat{\bold{H}}_k^{H} \bold{Y}_k$
at the output of the frequency-domain linear detector (see Fig. \ref{OFDM_MIMO_Block_1}).

It should be noted that the $jth$ component of $\bold{\widehat{H}}_k^{H} \bold{Y}_k$ is given by $\sum\limits_{i=1}^{N_R} \widehat{H}_k^{(i,j)*} Y_k^{(i)}$ $(j=1,2,\cdots, N_T)$: this means that the $\widehat{\textbf{H}}_k^H$ factor provides $N_T$ \MRC\ procedures, one per \MT\ antenna, all of them based on an appropriate \MF\ for each component of the length-$N_R$ received vector at subchannel $k$.

For an \MMSE\ detection -the optimum linear detection-, we need to adopt $\textbf{B}_k =  \textbf{A}_k^{-1}$, where 
\begin{eqnarray}\label{eq5}
\textbf{A}_k =  \widehat{\textbf{H}}_k^H\widehat{\textbf{H}}_k +\boldsymbol{\widehat{\alpha}},
\end{eqnarray}
with $\boldsymbol{\hat{\alpha}}$ denoting a diagonal matrix characterized by the estimated values of $\alpha^{(j)} =  \frac{N_o}{\left[\sigma_s^{(j)}\right]^2}\left(j=1,2,\dots N_T\right)$. By simply adopting $\textbf{B}_k=\textbf{I}_{N_T}$ (an $N_T\times N_T$ identity matrix), we can strongly reduce the implementation complexity: since $\textbf{D}_k = \textbf{B}_k  \widehat{\textbf{H}}_k^H = \widehat{\textbf{H}}_k$, this is a \MF\ detection.

When replacing the \MMSE\ detection by a simple \MF\ detection, however, a significant performance degradation is unavoidable. An intermediate detection solution - regarding both complexity and performance - is provided by the adoption of a simplified \MMSE-type scheme which replaces the exact inversion of $\textbf{A}_k$ by an approximate inversion. In this context, we consider a recommendation of \cite{ISCAS2013_wu} so as to obtain $\textbf{B}_k \approx \textbf{A}_k^{-1}$ while avoiding a complex $N_T\times N_T$ matrix inversion: $\textbf{B}_k = \left(diag\left(\textbf{A}_k\right)\right)^{-1} \times$ 
\begin{eqnarray}\label{eq6}
 \left[\textbf{I}_{N_T} - \left( \textbf{A}_k - diag\left(\textbf{A}_k\right) \right) \left(diag\left(\textbf{A}_k\right)\right)^{-1} \right],
\end{eqnarray}
where $\textbf{A}_k$ is given by (\ref{eq5}) and $diag\left(\textbf{A}_k\right)$ is a diagonal $N_T\times N_T$ matrix - very easy to invert - which shares the main diagonal with $\textbf{A}_k$.

By defining $\textbf{G}_k = \textbf{H}_k^H \textbf{H}_k$ - and $\widehat{\textbf{G}}_k = \widehat{\textbf{H}}_k^H \widehat{\textbf{H}}_k$ -, $\textbf{B}_k$ can be written as a function of $\widehat{\textbf{G}}_k$ and $\boldsymbol{\widehat{\alpha}}$ with both the standard \MMSE\ detection and the simplified \MMSE-type detection reported above, since $\textbf{A}_k = \widehat{\textbf{G}}_k + \boldsymbol{\widehat{\alpha}}$.

For a given channel realization $\bold{H}_k$ and a given detection matrix $\bold{D}_k$, which depends on the estimated channel realization $\bold{\widehat{H}}_k$, the output of the frequency-domain detector is given by
\begin{eqnarray}\label{eq9}
 \bold{\tilde{Y}}_k = \bold{D}_k \bold{Y}_k = \bold{\Gamma}_k \bold{S}_k + \bold{N'}_k,
\end{eqnarray}
where  $\bold{\Gamma}_k = \bold{D}_k \bold{H}_k $ and $\bold{N'}_k = \bold{D}_k \bold{N}_k$.

With \SC/\FDE\ (time-domain data symbols), an \IDFT\ is required for each  $\bold{\widetilde{Y}}^{(j)} = \left[\widetilde{Y}_0^{(j)},\widetilde{Y}_1^{(j)},\cdots,\widetilde{Y}_{N-1}^{(j)}\right]^T$ vector. The $n$th component of the resulting length-$N$ $IDFT\left(\bold{\widetilde{Y}}^{(j)} \right) = \bold{\widetilde{y}}^{(j)}$ vector can be written as $\widetilde{y}_n^{(j)} = \gamma^{(j)}s_n^{(j)} + ISI + MUI/MSI + \ 'Gaussian \ noise'$, with $\gamma^{(j)} = \frac{1}{N}\sum\limits_{k=0}^{N-1} \Gamma_k^{(j,j)}$ \hspace{1cm} $\left( \gamma^{(j)} = \frac{E\left[ \widetilde{y}_n^{(j)} s_n^{(j)*}\right] }{\left[\sigma_s^{(j)}\right]^2}\right) $.

Therefore, $\bold{\widetilde{Y}}_k$ can be written as
\begin{eqnarray}\label{eq11}
    \bold{\widetilde{Y}}_k  = \boldsymbol{\gamma} \bold{S}_k + \left(\bold{\Gamma}_k - \boldsymbol{\gamma}\right)\bold{S}_k + \bold{D}_k \bold{N}_k,
\end{eqnarray}
where $\boldsymbol{\gamma}$ is a diagonal $N_T\times N_T$ matrix with $(j,j)$ entries given by $\gamma^{(j)} = \frac{1}{N} \sum\limits_{k=0}^{N-1} \Gamma_k^{(j,j)}$. With $\widetilde{Y}_k^{(j)}$ written as
\begin{eqnarray}\label{eq12}
    \tilde{Y}_k^{(j)} = \gamma^{(j)} S_k^{(j)} + \left[\Gamma_k^{(j,j)} - \gamma^{(j)}\right]S_k^{(j)} + \\ \nonumber
    \begin{array}{c} \sum\limits_{l=1}^{N_T}\\*[-0.1cm]_{(l\neq j)}  \end{array} \Gamma_k^{(j,l)} S_k^{(l)} + \sum\limits_{i=1}^{N_R} D_k^{(j,i)} N_k^{(i)},
\end{eqnarray}
the terms in the right-hand side of eq. (\ref{eq12}) are concerned, respectively, to ''signal'', \ISI, \MUI/\MSI\ and ''Gaussian noise'', at subchannel $k$.

\subsection{Low-Complexity Iterative DF Detection Technique}\label{subsec2C}

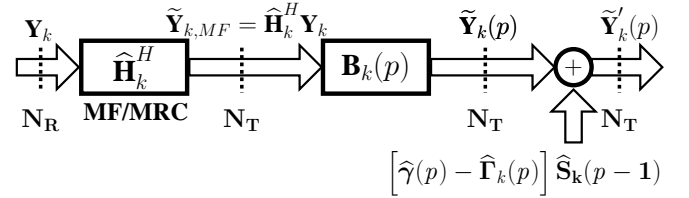
\begin{figure}[!ht]
	\centering
		
\tikzstyle{block} = [draw, fill=white, rectangle,minimum height=1.5cm, minimum width=3cm]	
\tikzstyle{sum} = [draw, circle, minimum size=1cm, node distance=1.75cm,line width=0.125cm]
\resizebox {\columnwidth} {!} {
\begin{tikzpicture}[
auto,
start chain = A going right,
every join/.style = {draw, -stealth, thick},
block/.append style = {on chain=A},
node distance=2.5cm,
font=\huge,
]
\node (a) at (0,0) {};
\node [label={[xshift=.75cm, yshift=0.5cm] $\textbf{Y}_{k}$}] {};
\node [single arrow,minimum size=1.25cm,draw=black,fill opacity=1,minimum height=5em, outer sep=0pt,line width=0.0625cm,right=0pt of a.east] {};
\node [block,line width=0.125cm,fill opacity=1,right of=input,node distance=.45cm,font=\Huge]  (b) at (3,0) {$\widehat{\textbf{H}}^H_{k}$};
\node [single arrow,minimum size=1.25cm,draw=black, fill opacity=1,minimum height=3.75cm,outer sep=0pt,line width=0.0625cm,right=0pt of A-1.east] {};
\node [label={[xshift=6.65cm, yshift=0.55cm] $\widetilde{\textbf{Y}}_{k,MF}=\widehat{\textbf{H}}^H_{k}\textbf{Y}_{k}$}] {};
\node [block, line width=0.125cm, fill opacity=1,right of=input,node distance=5.8cm,font=\Huge] (c) at (2,0) {$\textbf{B}_k (p)$};
\node [single arrow,minimum size=1.25cm,draw=black, fill opacity=1,minimum height=3.5cm, outer sep=0pt,line width=0.0625cm,right=0pt of A-2.east] {};
\node [label={[xshift=13.5cm, yshift=0.5cm] $\widetilde{\textbf{Y}}_{k} (p)$}] {};
\node [sum,minimum height=0.5cm] at (15.9,0) (sum) {{\huge $\bf{+}$}};
\node [single arrow,minimum size=1.25cm,draw=black, fill opacity=1,minimum height=2cm, outer sep=0pt,line width=0.0625cm] at (17.3,0) {};
\node [label={[xshift=13.5cm, yshift=0.5cm] $\widetilde{\textbf{Y}}_{k} (p)$}] {};
\node [label={[xshift=17.5cm, yshift=0.5cm] $\widetilde{\textbf{Y}}'_{k} (p)$}] {};

\node [single arrow,minimum size=1.25cm,rotate=+90,draw=black, fill opacity=1,minimum height=1.5cm, outer sep=0pt,line width=0.0625cm] at (15.9,-1.5) {};
\node [font=\huge] (d) at (14.5,-3.0) {$\left[ \boldsymbol{\widehat{{\gamma}}}(\it{p}) - \boldsymbol{\widehat{\Gamma}}_k(\it{p})\right] \bf{\widehat{S}}_k (\it{p}-\bf{1})$};
\node [font=\huge] (d) at (3.5,-1.25) {\bf MF/MRC};

\node[state,draw=none] (a1)  at (.8,1) {};
\node[state,draw=none] (a2)  at (6.5,1) {};
\node[state,draw=none] (a3)  at (13.4,1) {};
\node[state,draw=none] (a4)  at (17.2,1) {};
\begin{pgfinterruptboundingbox}
     \node[state,draw=none] (b1) [below of=a1] {$\bf{N_R}$};
     \node[state,draw=none] (b2) [below of=a2] {$\bf{N_T}$};
     \node[state,draw=none] (b3) [below of=a3] {$\bf{N_T}$};
     \node[state,draw=none] (b4) [below of=a4] {$\bf{N_T}$};
\end{pgfinterruptboundingbox}
\draw [-,dashed,line width=0.1cm] (a1) to[below]  node[auto] {} (b1);
\draw [-,dashed,line width=0.1cm] (a2) to[below]  node[auto] {} (b2);
\draw [-,dashed,line width=0.1cm] (a3) to[below]  node[auto] {} (b3);
\draw [-,dashed,line width=0.1cm] (a4) to[below]  node[auto] {} (b4);
\end{tikzpicture}
}\vspace{-.375cm}
	\caption{Iterative DF detection combining a linear detection and interference cancellation in the frequency domain.}   \label{OFDM_MIMO_Block_2}
\end{figure}

A low-complexity iterative \DF\ technique can be easily devised having in mind eq.  (\ref{eq11}). This frequency-domain non-linear detection technique is  depicted in Fig. \ref{OFDM_MIMO_Block_2}. It combines the use of a linear detector and, for all iterations after the initial iteration (i.e., for $p>1$), a cancellation of residual \MUI\ - and residual \MSI, when some users adopt several TX antennas for spatial multiplexing purposes - as well as residual \ISI; such cancellation is based on the estimated data block which is provided by the preceding iteration and fed back to the frequency-domain detector. The output of this detector, for iteration $p$, is
\begin{eqnarray}\label{eq13}
    \bold{\widetilde{Y}}'_k (p) = \bold{\widetilde{Y}}_k(p)+ \left[ \boldsymbol{\widehat{\gamma}}(p) - \bold{\widehat{\Gamma}}_k(p)\right] \bold{\widehat{S}}_k (p-1),
\end{eqnarray}
$\left[k=0,\cdots, N-1; p>1 \ (\text{for} \left. p=1,  \bold{\widetilde{Y}}'_k (p) = \bold{\widetilde{Y}}_k(p)\right) \right]$, where $\bold{\widehat{\Gamma}}_k (p) = \bold{D}_k (p) \bold{\widehat{H}}_k$ - with $\bold{D}_k (p)$ denoting the detection matrix employed in iteration $p$ - and the entries $(j,j)$ of the diagonal matrix $\boldsymbol{\widehat{\gamma}} (p)$ are given by $\widehat{\gamma}^{(j)} (p) = \frac{1}{N} \sum\limits_{k=0}^{N-1} \widehat{\Gamma}_k^{(j,j)} (p)$. Of course, $\left[\widehat{S}_0^{(j)}(p-1),\cdots,\widehat{S}_{N-1}^{(j)}(p-1)\right]^T = DFT\left(\left[\widehat{s}_0^{(j)}(p-1),\cdots,\widehat{s}_{N-1}^{(j)}(p-1)\right]^T\right)$.

The implementation of this iterative \DF\ technique is especially simple when $\bold{D}_k (p) = \bold{\widehat{H}}_k^H$ for any $p$, i.e., with a linear \MF\ detector for all iterations; an interesting alternative, which also avoids a complex matrix inversion, consists of adopting $\textbf{B}_k(p) = \textbf{I}_{N_T}$ for $p>1$ (i.e., \MF\ detection) and, for $p=1$, $\textbf{B}_k(p)$ as given by eq. (\ref{eq6}). It should be noted that $\bold{\widehat{\Gamma}}_k(p) = \bold{D}_k(p) \bold{\widehat{H}}_k = \bold{B}_k(p) \bold{\widehat{H}}_k^H \bold{\widehat{H}}_k = \bold{B}_k(p) \bold{\widehat{G}}_k$: $\bold{\widehat{\Gamma}}_k(p) = \bold{\widehat{G}}_k$ for the \MF\ detector and $\bold{\widehat{\Gamma}}_k(p)$ is a function of $\bold{\widehat{G}}_k$ and $\boldsymbol{\widehat{\alpha}}$ for the simplified \MMSE-type detector.

\section{On Achievable Detection Performances when Many BS Antennas are Available}\label{sec3}

\subsection{\BER\ with Linear Detection}\label{subsec3A}

Regarding evaluation of linear detection performances by simulation, a simple semi-analytical method is presented here, combines simulated channel realizations and analytical computations of \BER\ performance which are conditional on those channel realizations. In all cases, the conditional \BER\ values are directly computed by resorting to a \SINR, under the realistic assumption that the ''interference'' has a quasi-Gaussian nature. These ratios are simply derived in accordance with the channel realization $\bold{H}_k$ ($k=0,1,\cdots, N-1$). Of course, for concluding the \BER\ computation in each case - involving random generation of a large number of channel realizations and conditional \BER\ computations - a complementary averaging operation over the set of channel realizations is required.

When using a linear detection technique (sec. \ref{subsec2B}), it is easy to conclude, having in mind  \cite{GC2007_Revista2}, that the ''signal-to-interference-plus-noise'' ratio concerning the $j$th input of the MU-MIMO system is given by
\begin{eqnarray}\label{eq14}
 SINR_{j} =
&\frac{N\left|\bold{\gamma}^{(j)}\right|^2}{\beta_j + \begin{array}{c} \sum\limits_{l=1}^{N_T}\\*[-0.1cm]_{(l\neq j)}  \end{array}\beta_l  + \alpha^{(j)}\sum\limits_{i=1}^{N_R}\sum\limits_{k=0}^{N-1}\left|D_k^{(j,i)}\right|^2}
\end{eqnarray}
where $\alpha^{(j)} = \frac{N_0}{\left[\sigma_s^{(j)}\right]^2} = \frac{P_{\Sigma}}{2\eta m(j) \frac{E_b^{(j)}}{N_0}}$, $\beta_j = \sum\limits_{k=0}^{N-1}\left|\Gamma_k^{(j,j)}-\gamma^{(j)}\right|^2$,  $\beta_l = \frac{\left[\sigma_s^{(l)}\right]^2}{\left[\sigma_s^{(j)}\right]^2}\sum\limits_{k=0}^{N-1}\left|\Gamma_k^{(j,l)}\right|^2 \ (l\neq j)$.

For $4$-\QAM\ transmission ($m(j)=1$), the resulting $BER_j$ ($j=1,2,\cdots, N_T$) - conditional on the channel realization $\{\bold{H}_k; \ k = 0,1,\cdots,N-1\}$ - is given by 
\begin{eqnarray}\label{eq15}
    BER_j \approx Q\left(\sqrt{SINR_j}\right),
\end{eqnarray}
where $Q(.)$ is the Gaussian error function) with $SINR_j$ as computed above.

For $16$-\QAM\ transmission ($m(j)=2$), the resulting $BER_j$ ($j=1,2,\cdots, N_T$) can be obtained  very accurately as follows (having in mind the Gray mapping rule):
\begin{eqnarray}\label{eq16}
    BER_j = \frac{1}{2} \left[BER_{j,1} + BER_{j,2}\right]
\end{eqnarray}
where $BER_{j,1} =  \frac{1}{2} p_{1} + \frac{1}{2} p_{3}$ and $BER_{j,2} =  \frac{1}{2} \left(p_{1} + p_{3}\right) + \frac{1}{2} \left(p_{1} - p_{5}\right) $ with 
\begin{eqnarray}\label{eq17}
    p_n \approx Q \left(\sqrt{\frac{n^2}{5} SINR_j}\right).
\end{eqnarray}

For $64$-\QAM\ transmission ($m(j)=3$), an accurate approximation to $BER_j$ can also be obtained  (also having in mind the Gray mapping rule):
\begin{eqnarray}\label{eq18}
    BER_j = \frac{1}{3} \left[BER_{j,1} + BER_{j,2} + BER_{j,3}\right]
\end{eqnarray}
where $BER_{j,1} =  \frac{1}{4} p'_{1} + \frac{1}{4} p'_{3} + \frac{1}{4} p'_{5} + \frac{1}{4} p'_{7}$, $BER_{j,2} =  \frac{1}{4} \left(p'_{3} - p'_{11}\right) + \frac{1}{4} \left(p'_{1} - p'_{9}\right) + \frac{1}{4} \left(p'_{1} + p'_{7}\right)  + \frac{1}{4} \left(p'_{3} + p'_{5}\right)$ and  $BER_{j,3} =  \frac{1}{4} \left(p'_{1} - p'_{5} + p'_{9} - p'_{13}\right) + \frac{1}{4} \left(p'_{1} + p'_{3} - p'_{7} + p'_{11}\right) + \frac{1}{4} \left(p'_{3} + p'_{1} - p'_{5} + p'_{9}\right) + \frac{1}{4} \left(p'_{1} - p'_{5} + p'_{3} - p'_{7}\right)$,
with
\begin{eqnarray}\label{eq19}
    p'_n \approx Q \left(\sqrt{\frac{n^2}{21} SINR_j}\right).
\end{eqnarray}

When $n>1$, of course, $p_n \ll p_1$ and $p'_n \ll p'_1$ for high $SINR_j$. By using $M(j) = 2^{m(j)}$, we get
\begin{eqnarray}\label{eq20}
    BER_j \approx \frac{2}{m(j)}\left(1-\frac{1}{M(j)}\right)  Q \left(\sqrt{3\frac{ SINR_j}{\left[M(j)\right]^2-1} }\right)
\end{eqnarray}

\subsection{Irreducible \BER\ Levels under Low-Complexity Implementation Constraints}\label{subsec3B}

When a low-complexity linear detection technique is adopted - such as the \MF\ detection or the simplified \MMSE-type detection reported in subsection \ref{subsec2B} -, an "error floor" effect is unavoidable. The corresponding Irreducible \BER\ is given by $IBER_j = \lim\limits_{\alpha^{(j)}\rightarrow 0} BER_j =$
\begin{eqnarray}\label{eq21}
    \lim\limits_{\alpha^{(j)}\rightarrow 0} f\left(SINR_j\right) = f\left(\frac{N\left|\gamma^{(j)}\right|^2}{\beta_j +\sum\limits_{\begin{array}{c}\vspace{-.125cm} l=1 \ (l\neq j) \end{array} }^{N_T} \beta_l}\right) 
\end{eqnarray}
where $ f\left(SINR_j\right)$ is the appropriate function of $ SINR_j$, according to $m(j)$, which is used for computation of $BER_j$, as explained in Sub-Section \ref{subsec3A} $\left(\gamma^{(j)}, \beta_j \ \text{and} \ \beta_l \ (l\neq j) \ \text{as in } (\ref{eq14}) \right)$.

\subsection{\SIMO\ Performances Bounds and Massive \MIMO\ effects}\label{subsec3C}

The $SINR_j$ for semi-analytical computation of the \SIMO/\MFB\ on detection performance can be easily derived from (\ref{eq14}), by considering the Gaussian noise and the ''useful'' term, at the \MF\ detector output, but not the two interference terms (\MUI/\MSI, \ISI). Therefore, the \SIMO/\MFB\ bound is given by the appropriate $f\left(SINR_j\right)$, according to $m(j)$, with $SINR_j = $
\begin{equation}
\frac{N\left|\gamma^{(j)}\right|^2}{\alpha^{(j)} \sum\limits_{k=0 }^{N-1}\sum\limits_{i=1 }^{N_R} \left|H_k^{(i,j)}\right|^2} =  \frac{1}{N\alpha^{(j)}}\sum\limits_{k=0 }^{N-1}\sum\limits_{i=1 }^{N_R} \left|H_k^{(i,j)}\right|^2
\end{equation}
since $\gamma^{(j)} = \frac{1}{N}\sum\limits_{k=0 }^{N-1}  \sum\limits_{i=1 }^{N_R} \left|H_k^{(i,j)}\right|^2$ with \MF\ detection.

By assuming a \LOS\ single-path radio propagation for each ($i,j$) antenna pair - i.e., $\left|H_k^{(i,j)}\right|^2 = P_\Sigma$ -, we can obtain
\begin{equation}
SINR_j = \frac{1}{\alpha^{(j)}} N_R P_\Sigma = 2\eta m(j) N_R \frac{E_b^{(j)}}{N_0}
\end{equation}
leading to the \SIMO/AWGN/\MFB.

When $N_R \gg N_T$, both the \MUI/\MSI\ effects and the effects of multipath propagation (fading, \ISI) tend to disappear: consequently, the \BER\ performances for the \MU-\MIMO\ $N_T \times N_R$ Rayleigh fading channel become very close  to those concerning a \SIMO\  $1\times N_R$ channel with single-path propagation for all $N_R$ TX/RX antenna pairs.  The "massive MIMO" effects that can be explained, when $N_R \gg N_T$, mean that $\sum\limits_{i=1 }^{N_R} \left|H_k^{(i,j)}\right|^2 \approx N_R P_\Sigma$ and $\sum\limits_{\begin{array}{c}\vspace{-.125cm} l=1 \ (l\neq j) \end{array} }^{N_T} H_k^{(i,j)*}H_k^{(i,l)}\approx 0$, leading to $BER_j = f\left(SINR_j\right)$ with $SINR_j \approx 2\eta m(j) N_R \frac{E_b^{(j)}}{N_0}$ (i.e., closely approximating the \SIMO/AWGN/\MFB).

\section{Numerical Results and Discussion}\label{sec4}

The set of performance results which are presented here are concerned to \SC/\FDE\ uplink block transmission using \QAM\ (up to $6$ bit/symbol), with $N=256$ and $Ls=64$ in a \MU-\MIMO\ $N_T\times N_R$ Rayleigh fading channel with $N_T = 12$ or $N_T = 24$. Perfect channel estimation and power control are assumed. The fading effects regarding the several TX/RX antenna pairs are supposed to be uncorrelated, with independent zero-mean complex Gaussian $h_n^{(i,j)}$ coefficients assumed to have variances $P_n = 1 - \frac{n}{63}$, $n=0,...,63$ ($P_n=0$ for $n=64,...,255$).

\begin{figure}
	\begin{minipage}{.5\textwidth}
			
\pgfplotstableread{./error_floor/error_floor_PIC_Nt12.dat}\myerrorfloor

 	\begin{tikzpicture} 
 	
 	\begin{semilogyaxis}[
 	clip=true, 
 	font=\boldmath,
 	ultra thick,
 	legend pos=north east,
 	xmin=20,xmax=200,ymin=1e-3,ymax=1e-0,ymode=log,grid=both,
 	height = 6cm,width=8.5cm,
 	xtick  = {20,40,...,200},ytickten={-3,...,0},yticklabel pos=left,
 	xlabel = {$\tiny\bf N_R$},	
 	ylabel = \small\bf IBER,
 	grid   = none,
	legend entries = {\tiny\bf MMSE-type,\tiny\bf MF},
	legend style={at={(1,1)},anchor=north east,draw = none,fill=none},
	legend cell align=left,
 	]
 	\tikzstyle{fontfiglisas} = [smooth,color=black,line width=1.75pt,mark options={solid,scale=1,fill=black}]
	\addplot+[fontfiglisas,mark=none]  
	table [x={NR} , y=ber_MMSE_simples_m1,col sep=space] {\myerrorfloor};	
	
	\addplot+[fontfiglisas,mark=*,forget plot]  
	table [x={NR} , y=ber_MMSE_simples_m1, col sep=space] {\myerrorfloor};	
	
	\addplot+[fontfiglisas,mark=triangle*,forget plot]  
	table [x={NR} , y=ber_MMSE_simples_m2, col sep=space] {\myerrorfloor};	
	
	\addplot+[fontfiglisas,mark=diamond*,forget plot]  
	table [x={NR} , y=ber_MMSE_simples_m3, col sep=space] {\myerrorfloor};	
	\addplot+[fontfiglisas,mark=none,dashed]  
 	table [x={NR} , y=ber_MF_m1, col sep=space] {\myerrorfloor};	

	\addplot+[fontfiglisas,mark=*,dashed,forget plot]  
 	table [x={NR} , y=ber_MF_m1, col sep=space] {\myerrorfloor};	
 	
	\addplot+[fontfiglisas,mark=triangle*,dashed,forget plot]  
 	table [x={NR} , y=ber_MF_m2, col sep=space] {\myerrorfloor};	

	\addplot+[fontfiglisas,mark=diamond*,dashed,forget plot]  
	table [x={NR} , y=ber_MF_m3, col sep=space] {\myerrorfloor};	


  	\end{semilogyaxis}
 	\end{tikzpicture}
		\vspace{-.1cm}\begin{quote} \hspace{1.0cm}\centering{\bf(a)} \end{quote}
	\end{minipage}
	\begin{minipage}{.5\textwidth}\vspace{-.00cm}
			
\pgfplotstableread{./error_floor/error_floor_PIC_Nt24.dat}\myerrorfloor
	

 	\begin{tikzpicture} 

 	\begin{semilogyaxis}[
 	clip=true, 
 	font=\boldmath,
 	ultra thick,
 	legend pos=north east,
 	xmin=40,xmax=400,ymin=1e-3,ymax=1e-0,ymode=log,grid=both,
 	height = 6cm,width=8.5cm,
 	xtick  = {40,80,...,400},ytickten={-3,...,0},yticklabel pos=left,
 	xlabel = {$\tiny\bf N_R$},	
 	ylabel = \small\bf IBER,
 	grid   = none,
	legend entries = {\tiny\bf MMSE-type,\tiny\bf MF},
	legend style={at={(1,1)},anchor=north east,draw = none,fill=none},
	legend cell align=left,
 	]
 	\tikzstyle{fontfiglisas} = [smooth,color=black,line width=1.75pt,mark options={solid,scale=1.0,fill=black}]
	\addplot+[fontfiglisas,mark=none]  
	table [x={NR} , y=ber_MMSE_simples_m1,col sep=space] {\myerrorfloor};	
	
	\addplot+[fontfiglisas,mark=*,forget plot]  
	table [x={NR} , y=ber_MMSE_simples_m1, col sep=space] {\myerrorfloor};	
	
	\addplot+[fontfiglisas,mark=triangle*,forget plot]  
	table [x={NR} , y=ber_MMSE_simples_m2, col sep=space] {\myerrorfloor};	
	
	\addplot+[fontfiglisas,mark=diamond*,forget plot]  
	table [x={NR} , y=ber_MMSE_simples_m3, col sep=space] {\myerrorfloor};	
	\addplot+[fontfiglisas,mark=none,dashed]  
 	table [x={NR} , y=ber_MF_m1, col sep=space] {\myerrorfloor};	

	\addplot+[fontfiglisas,mark=*,dashed,forget plot]  
 	table [x={NR} , y=ber_MF_m1, col sep=space] {\myerrorfloor};	
 	
	\addplot+[fontfiglisas,mark=triangle*,dashed,forget plot]  
 	table [x={NR} , y=ber_MF_m2, col sep=space] {\myerrorfloor};	

	\addplot+[fontfiglisas,mark=diamond*,dashed,forget plot]  
	table [x={NR} , y=ber_MF_m3, col sep=space] {\myerrorfloor};	

  	\end{semilogyaxis}
 	\end{tikzpicture}
		\vspace{-.1cm}\begin{quote} \hspace{1.0cm}\centering{\bf(b)} \end{quote}
	\end{minipage}
	\caption{IBER performance curves concerning either MF detection (dashed lines) or simplified \MMSE-type detection (solid lines), when $N_T=12$ (a) or $N_T=24$ (b) for $4-$ ($\bullet$) (the best), $16-$ ($\blacktriangle$) or $64-$QAM ($\blacklozenge$) (the worst).}  \label{Fig:simsSC_SIR_error_floor}
\end{figure}
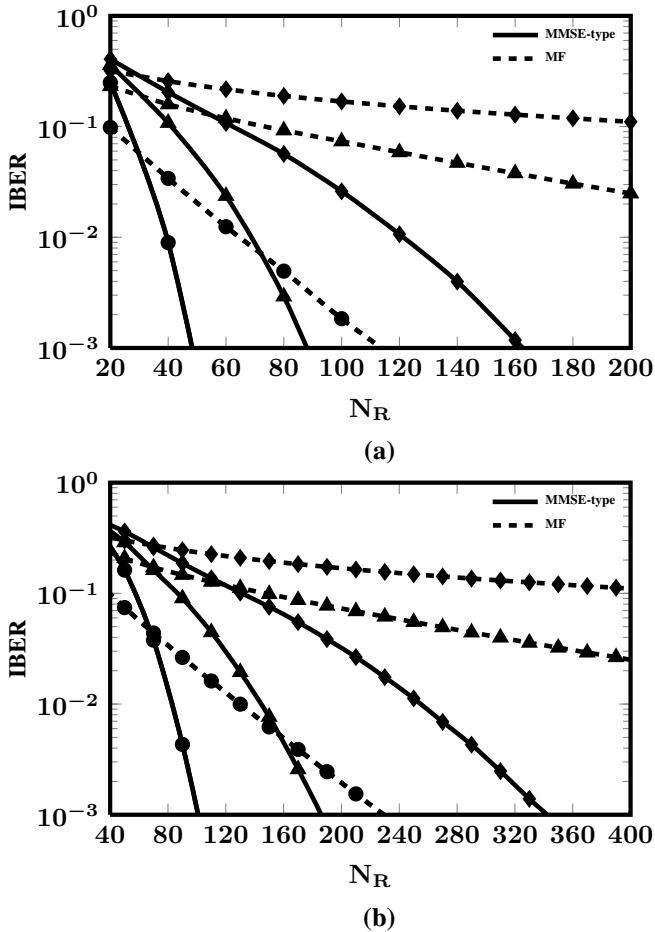

Fig. \ref{Fig:simsSC_SIR_error_floor} shows \IBER\ results for a wide range of values of $N_R$  and the same \QAM\ constellation ($4$, $16$ or $64-$\QAM) at all $N_T$ inputs of the \MU-\MIMO\ system. These performance curves have been obtained through the semi-analytical simulation approach presented in sec. \ref{sec3}, and confirmed by conventional Monte-Carlo simulation (quasi-superposed 
dots), involving an error-counting procedure. Two alternative low-complexity linear detection techniques were considered, both avoiding complex $N_T \times N_T$ matrix inversions: the \MF\ detection and the simplified \MMSE-type detection, as reported in sec. \ref{sec2}. These results clearly show that, under both low-complexity detection techniques, unacceptably high error floor levels (say, $\left.IBER>10^{-1}\right)$ are unavoidable if $N_R$ is not large enough; they become especially high when a high bandwidth efficiency is intended - through an increased size of the \QAM\ constellation - and/or the \MF\ detection is adopted. An excellent agreement of performance results obtained by semi-analytical means and by conventional Monte Carlo simulations is also observed. By comparing Fig. \ref{Fig:simsSC_SIR_error_floor}(a) and Fig. \ref{Fig:simsSC_SIR_error_floor}(b)  ( $N_T = 12$ and   $N_T = 24$, respectively), we can notice that the IBER level is essentially determined by the $N_R/N_T$  ratio, for each \QAM\ scheme and each linear detection technique; we can also notice that the simplified \MMSE-type detection outperforms the \MF\ detection when that ratio is above a given threshold ($N_R/N_T$ approximately equal to $2.5$ in all cases).

  \begin{figure}
  	\begin{minipage}{.5\textwidth}
  			
\pgfplotstableread{./my_figs3/Nt12_Nr60_MF_MMSE_MMSEsimples_m1.dat}\mySIMSmum
\pgfplotstableread{./my_figs3/Nt12_Nr60_MF_MMSE_MMSEsimples_m2.dat}\mySIMSmdois
\pgfplotstableread{./my_figs3/Nt12_Nr60_MF_MMSE_MMSEsimples_m3.dat}\mySIMSmtres
 	
 	
 	\begin{tikzpicture} 
 	
 	\begin{semilogyaxis}[
 	font=\boldmath,
 	ultra thick,
 	clip=true, 
 	xmin=-20,xmax=4,ymin=1e-4,ymax=1e-0,ymode=log,grid=both,
 	height = 7cm,
 	xtick  = {-20,-12,...,6},ytickten={-7,...,0},yticklabel pos=left,
 	xlabel = {$\tiny\bf E_b/N_0 (dB)$},	
 	ylabel = \small\bf Bit Error Rate,
 	grid   = none,
 	legend entries = {\tiny\bf MF,\tiny\bf S.MMSE,\tiny\bf MMSE,\tiny\bf Bound},
 	legend style={at={(0,0)},anchor=south west,draw = none,fill=none},
 	legend cell align=right,
 	]
	\tikzstyle{fontfiglisas} = [smooth,color=black,line width=1.0pt,mark options={solid,scale=1.25,fill=white}]
 	\addplot+[fontfiglisas,color=black,dashed,line width=2pt,mark=none]  
 	table [x={SNR} , y=ber_MF, col sep=space] {\mySIMSmum};
	
 	\addplot+[fontfiglisas,color=black,solid,line width=2pt,mark=none]  
 	table [x=SNR, y=ber_MMSE_simples, col sep=space] {\mySIMSmum};

 	\addplot+[fontfiglisas,color=black,dashdotted,line width=2pt,mark=none]  
 	table [x=SNR , y=ber_MMSE, col sep=space] {\mySIMSmum};
 	
 	\addplot+[fontfiglisas,color=red,dotted,line width=2pt,mark=none]  
 	table [x=SNR, y=ber_theory, col sep=space] {\mySIMSmum};
  	\addplot+[fontfiglisas,color=black,dashed,line width=2pt,mark=none,forget plot]  
  	table [x={SNR} , y=ber_MF, col sep=space] {\mySIMSmdois};
  	
  	\addplot+[fontfiglisas,color=black,solid,line width=2pt,mark=none,forget plot]  
  	table [x=SNR, y=ber_MMSE_simples, col sep=space] {\mySIMSmdois};
  	
  	\addplot+[fontfiglisas,color=black,dashdotted,line width=2pt,mark=none,forget plot]  
  	table [x=SNR , y=ber_MMSE, col sep=space] {\mySIMSmdois};
  	
  	\addplot+[fontfiglisas,color=red,dotted,line width=2pt,mark=none,forget plot]  
  	table [x=SNR, y=ber_theory, col sep=space] {\mySIMSmdois};
   	\addplot+[fontfiglisas,color=black,dashed,line width=2pt,mark=none,forget plot]  
   	table [x={SNR} , y=ber_MF, col sep=space] {\mySIMSmtres};
   	
   	\addplot+[fontfiglisas,color=black,solid,line width=2pt,mark=none,forget plot]  
   	table [x=SNR, y=ber_MMSE_simples, col sep=space] {\mySIMSmtres};
   	
   	\addplot+[fontfiglisas,color=black,dashdotted,line width=2pt,mark=none,forget plot]  
   	table [x=SNR , y=ber_MMSE, col sep=space] {\mySIMSmtres};
   	
   	\addplot+[fontfiglisas,color=red,dotted,line width=2pt,mark=none,forget plot]  
   	table [x=SNR, y=ber_theory, col sep=space] {\mySIMSmtres};	 		
 	\end{semilogyaxis}
 	\end{tikzpicture}
  		\vspace{-.25cm}\begin{quote} \hspace{0cm}\centering{\bf(a)} \end{quote}
  	\end{minipage}
  	\begin{minipage}{.5\textwidth}
  			
\pgfplotstableread{./my_figs3/Nt24_Nr120_MF_MMSE_MMSEsimples_m1.dat}\mySIMSmum
\pgfplotstableread{./my_figs3/Nt24_Nr120_MF_MMSE_MMSEsimples_m2.dat}\mySIMSmdois
\pgfplotstableread{./my_figs3/Nt24_Nr120_MF_MMSE_MMSEsimples_m3.dat}\mySIMSmtres
 	
 	
 	\begin{tikzpicture} 
 	
 	\begin{semilogyaxis}[
 	font=\boldmath,
 	ultra thick,
 	clip=true, 
 	xmin=-20,xmax=4,ymin=1e-4,ymax=1e-0,ymode=log,grid=both,
 	height = 7cm,
 	xtick  = {-20,-12,...,6},ytickten={-7,...,0},yticklabel pos=left,
 	xlabel = {$\tiny\bf E_b/N_0 (dB)$},	
 	ylabel = \small\bf Bit Error Rate,
 	grid   = none,
 	legend entries = {\tiny\bf MF,\tiny\bf S.MMSE,\tiny\bf MMSE,\tiny\bf Bound},
 	legend style={at={(0,0)},anchor=south west,draw = none,fill=none},
 	legend cell align=right,
 	]
	\tikzstyle{fontfiglisas} = [smooth,color=black,line width=1.0pt,mark options={solid,scale=1.25,fill=white}]
 	\addplot+[fontfiglisas,color=black,dashed,line width=2pt,mark=none]  
 	table [x={SNR} , y=ber_MF, col sep=space] {\mySIMSmum};
	
 	\addplot+[fontfiglisas,color=black,solid,line width=2pt,mark=none]  
 	table [x=SNR, y=ber_MMSE_simples, col sep=space] {\mySIMSmum};

 	\addplot+[fontfiglisas,color=black,dashdotted,line width=2pt,mark=none]  
 	table [x=SNR , y=ber_MMSE, col sep=space] {\mySIMSmum};
 	
 	\addplot+[fontfiglisas,color=red,dotted,line width=2pt,mark=none]  
 	table [x=SNR, y=ber_theory, col sep=space] {\mySIMSmum};
  	\addplot+[fontfiglisas,color=black,dashed,line width=2pt,mark=none,forget plot]  
  	table [x={SNR} , y=ber_MF, col sep=space] {\mySIMSmdois};
  	
  	\addplot+[fontfiglisas,color=black,solid,line width=2pt,mark=none,forget plot]  
  	table [x=SNR, y=ber_MMSE_simples, col sep=space] {\mySIMSmdois};
  	
  	\addplot+[fontfiglisas,color=black,dashdotted,line width=2pt,mark=none,forget plot]  
  	table [x=SNR , y=ber_MMSE, col sep=space] {\mySIMSmdois};
  	
  	\addplot+[fontfiglisas,color=red,dotted,line width=2pt,mark=none,forget plot]  
  	table [x=SNR, y=ber_theory, col sep=space] {\mySIMSmdois};
   	\addplot+[fontfiglisas,color=black,dashed,line width=2pt,mark=none,forget plot]  
   	table [x={SNR} , y=ber_MF, col sep=space] {\mySIMSmtres};
   	
   	\addplot+[fontfiglisas,color=black,solid,line width=2pt,mark=none,forget plot]  
   	table [x=SNR, y=ber_MMSE_simples, col sep=space] {\mySIMSmtres};
   	
   	\addplot+[fontfiglisas,color=black,dashdotted,line width=2pt,mark=none,forget plot]  
   	table [x=SNR , y=ber_MMSE, col sep=space] {\mySIMSmtres};
   	
   	\addplot+[fontfiglisas,color=red,dotted,line width=2pt,mark=none,forget plot]  
   	table [x=SNR, y=ber_theory, col sep=space] {\mySIMSmtres};	 		
 	\end{semilogyaxis}
 	\end{tikzpicture}
  		\vspace{-.25cm}\begin{quote} \hspace{0cm}\centering{\bf(b)} \end{quote}
  	\end{minipage}
  	\begin{minipage}{.5\textwidth}
  			
\pgfplotstableread{./my_figs3/Nt12_Nr120_MF_MMSE_MMSEsimples_m1.dat}\mySIMSmum
\pgfplotstableread{./my_figs3/Nt12_Nr120_MF_MMSE_MMSEsimples_m2.dat}\mySIMSmdois
\pgfplotstableread{./my_figs3/Nt12_Nr120_MF_MMSE_MMSEsimples_m3.dat}\mySIMSmtres
 	
 	
 	\begin{tikzpicture} 
 	
 	\begin{semilogyaxis}[
 	font=\boldmath,
 	ultra thick,
 	clip=true, 
 	xmin=-20,xmax=4,ymin=1e-4,ymax=1e-0,ymode=log,grid=both,
 	height = 7cm,
 	xtick  = {-20,-12,...,6},ytickten={-7,...,0},yticklabel pos=left,
 	xlabel = {$\tiny\bf E_b/N_0 (dB)$},	
 	ylabel = \small\bf Bit Error Rate,
 	grid   = none,
 	legend entries = {\tiny\bf MF,\tiny\bf S.MMSE,\tiny\bf MMSE,\tiny\bf Bound},
 	legend style={at={(0,0)},anchor=south west,draw = none,fill=none},
 	legend cell align=right,
 	]
	\tikzstyle{fontfiglisas} = [smooth,color=black,line width=1.0pt,mark options={solid,scale=1.25,fill=white}]
 	\addplot+[fontfiglisas,color=black,dashed,line width=2pt,mark=none]  
 	table [x={SNR} , y=ber_MF, col sep=space] {\mySIMSmum};
	
 	\addplot+[fontfiglisas,color=black,solid,line width=2pt,mark=none]  
 	table [x=SNR, y=ber_MMSE_simples, col sep=space] {\mySIMSmum};

 	\addplot+[fontfiglisas,color=black,dashdotted,line width=2pt,mark=none]  
 	table [x=SNR , y=ber_MMSE, col sep=space] {\mySIMSmum};
 	
 	\addplot+[fontfiglisas,color=red,dotted,line width=2pt,mark=none]  
 	table [x=SNR, y=ber_theory, col sep=space] {\mySIMSmum};
  	\addplot+[fontfiglisas,color=black,dashed,line width=2pt,mark=none,forget plot]  
  	table [x={SNR} , y=ber_MF, col sep=space] {\mySIMSmdois};
  	
  	\addplot+[fontfiglisas,color=black,solid,line width=2pt,mark=none,forget plot]  
  	table [x=SNR, y=ber_MMSE_simples, col sep=space] {\mySIMSmdois};
  	
  	\addplot+[fontfiglisas,color=black,dashdotted,line width=2pt,mark=none,forget plot]  
  	table [x=SNR , y=ber_MMSE, col sep=space] {\mySIMSmdois};
  	
  	\addplot+[fontfiglisas,color=red,dotted,line width=2pt,mark=none,forget plot]  
  	table [x=SNR, y=ber_theory, col sep=space] {\mySIMSmdois};
   	\addplot+[fontfiglisas,color=black,dashed,line width=2pt,mark=none,forget plot]  
   	table [x={SNR} , y=ber_MF, col sep=space] {\mySIMSmtres};
   	
   	\addplot+[fontfiglisas,color=black,solid,line width=2pt,mark=none,forget plot]  
   	table [x=SNR, y=ber_MMSE_simples, col sep=space] {\mySIMSmtres};
   	
   	\addplot+[fontfiglisas,color=black,dashdotted,line width=2pt,mark=none,forget plot]  
   	table [x=SNR , y=ber_MMSE, col sep=space] {\mySIMSmtres};
   	
   	\addplot+[fontfiglisas,color=red,dotted,line width=2pt,mark=none,forget plot]  
   	table [x=SNR, y=ber_theory, col sep=space] {\mySIMSmtres};	 		
 	\end{semilogyaxis}
 	\end{tikzpicture}
  		\vspace{-.25cm}\begin{quote} \hspace{0cm}\centering{\bf(c)} \end{quote}
  	\end{minipage}
  	\caption{BER performances - when $N_T = 12$ and $N_R = 60$ (a), $N_T = 24$ and $N_R = 120$ (b) or $N_T = 12$ and $N_R = 120$ (c) - concerning \MF\ detection (dashed lines), simplified \MMSE-type detection (solid lines) or true \MMSE\ detection (dot-dashed lines), for  $4$-\QAM\ (the best), $16$-\QAM\ or $64$-\QAM\ (the worst) in all TX antennas [SIMO/AWGN/MFB reference performances (dotted lines) are also included.]}  \label{Fig:simsSC_PIC_MF_ou_MMSEsimples}
  \end{figure}

Fig. \ref{Fig:simsSC_PIC_MF_ou_MMSEsimples} shows \BER\ performances - obtained through the semi-analytical 
method - with $N_T = 12$ and $N_R = 60$ (a),  $N_T = 24$ and $N_R = 120$ (b) or $N_T = 12$ and $N_R = 120$ (c), the same \QAM\ constellation and the same $\sigma_s^{(j)}$ at all $N_T$ system inputs, and a linear detection technique: \MF\ detection, simplified \MMSE-type detection or true MMSE detection (see subsection \ref{subsec2B}). For comparisons, we also include the analytical \SIMO/AWGN/\MFB, with the quasi-superposed dots denoting the \SIMO/\MFB, semi-analytically obtained. These results confirm those of Fig. \ref{Fig:simsSC_SIR_error_floor} on the performance limitations of the low-complexity linear detection techniques (designed to avoid a complex $N_T \times N_T$ matrix inversion), also showing the good performances achievable through a linear ideal \MMSE\ detection (which requires a complex $N_T \times N_T$ matrix inversion), with a nice approximation to the \SIMO/AWGN/\MFB\ (a gap below $2$ dB, even for $64-$QAM and the lower $N_R$). By comparing Fig. \ref{Fig:simsSC_PIC_MF_ou_MMSEsimples}(a) ($N_T = 12$,  $N_R = 60$) and Fig. \ref{Fig:simsSC_PIC_MF_ou_MMSEsimples}(b) ($N_T = 24$,  $N_R = 120$) - concerning the same ratio $N_R/N_T=5$, two times above the threshold - we can notice that, as expected, the achievable \BER\ for very high signal-to-noise ratios closely approximates the corresponding \IBER, shown in Fig. \ref{Fig:simsSC_SIR_error_floor}; by comparing both figures with Fig.  \ref{Fig:simsSC_PIC_MF_ou_MMSEsimples}(c) - concerning  $N_R/N_T=10$ -, we can observe the performance benefits of the higher ratio.

Fig. \ref{Fig:simsSC_PIC_MMSEsimplesMF} shows the simulated \BER\ performances - obtained through the conventional Monte Carlo method - with $N_T = 12$ and $N_R = 60$ (a), $N_T = 24$  and $N_R = 120$ (b) or $N_T = 12$  and $N_R = 120$ (c), the same \QAM\ constellation and the same $\sigma_s^{(j)}$ used at all $N_T$ system inputs, and an iterative \DF\ detection technique as described in subsection \ref{subsec2C} . A simplified \MMSE-type detector is adopted for the first iteration (i.e. $\textbf{B}_k (p)$ is given by eq. (\ref{eq6}) for $p = 1$) and an \MF\ detector is adopted for the remaining iterations (i.e. $\textbf{B}_k (p)= \textbf{I}_{N_T}$  for $p > 1$). Similarly to Fig. \ref{Fig:simsSC_PIC_MF_ou_MMSEsimples}, this figure also includes analytical \SIMO/AWGN/\MFB\ results. In the closely related Fig. \ref{Fig:simsSC_PIC_MF}, also concerning an iterative DF detection as described in subsection \ref{subsec2C}, we adopted an MF detector for all iterations.

\begin{figure}
\begin{minipage}{.5\textwidth}
		
\pgfplotstableread{./my_figs4/Nt12Nr60_it1MMSEsimples_MF_m1.dat}\mySIMSmum
\pgfplotstableread{./my_figs4/Nt12Nr60_it1MMSEsimples_MF_m2.dat}\mySIMSmdois
\pgfplotstableread{./my_figs4/Nt12Nr60_it1MMSEsimples_MF_m3.dat}\mySIMSmtres
 	
 	
 	\begin{tikzpicture} 
 	
 	\begin{semilogyaxis}[
 	font=\boldmath,
 	ultra thick,
 	clip=true, 
 	xmin=-20,xmax=4,ymin=1e-4,ymax=1e-0,ymode=log,grid=both,
 	height = 7cm,
 	xtick  = {-20,-12,...,6},ytickten={-7,...,0},yticklabel pos=left,
 	xlabel = {$\tiny\bf E_b/N_0 (dB)$},	
 	ylabel = \small\bf Bit Error Rate,
 	grid   = none,
 	legend entries = {\bf Bound,\bf 4QAM,\bf 16QAM,\bf 64QAM},
 	legend style={at={(0,0)},anchor=south west,draw = none,fill=none,font=\fontsize{6}{5}\selectfont},
 	legend cell align=right,
 	]
	\tikzstyle{fontfiglisas} = [smooth,color=black,line width=1.0pt,mark options={solid,scale=1.25,fill=white}]
 	\addplot+[fontfiglisas,color=red,dotted,line width=2pt,mark=none]  
 	table [x={SNR} , y=ber_theory, col sep=space] {\mySIMSmum};
	
 	\addplot+[fontfiglisas,color=black,solid,line width=2pt,mark=none]  
 	table [x=SNR, y=ber_sims1, col sep=space] {\mySIMSmum};

 	\addplot+[fontfiglisas,color=black,solid,line width=2pt,mark=none,forget plot]  
 	table [x=SNR , y=ber_sims2, col sep=space] {\mySIMSmum};
 	
 	\addplot+[fontfiglisas,color=black,solid,line width=2pt,mark=none,forget plot]  
 	table [x=SNR, y=ber_sims3, col sep=space] {\mySIMSmum};
  	\addplot+[fontfiglisas,color=red,dotted,line width=2pt,mark=none,forget plot]  
  	table [x={SNR} , y=ber_theory, col sep=space] {\mySIMSmdois};
  	
  	\addplot+[fontfiglisas,color=black,dashed,line width=2pt,mark=none]  
  	table [x=SNR, y=ber_sims1, col sep=space] {\mySIMSmdois};
  	
  	\addplot+[fontfiglisas,color=black,dashed,line width=2pt,mark=none,forget plot]  
  	table [x=SNR , y=ber_sims2, col sep=space] {\mySIMSmdois};
  	
  	\addplot+[fontfiglisas,color=black,dashed,line width=2pt,mark=none,forget plot]  
  	table [x=SNR, y=ber_sims3, col sep=space] {\mySIMSmdois};
   	\addplot+[fontfiglisas,color=red,dotted,line width=2pt,mark=none,forget plot]  
   	table [x={SNR} , y=ber_theory, col sep=space] {\mySIMSmtres};
   	
   	\addplot+[fontfiglisas,color=black,dashdotted,line width=2pt,mark=none]  
   	table [x=SNR, y=ber_sims1, col sep=space] {\mySIMSmtres};
   	
   	\addplot+[fontfiglisas,color=black,dashdotted,line width=2pt,mark=none,forget plot]  
   	table [x=SNR , y=ber_sims2, col sep=space] {\mySIMSmtres};
   	
   	\addplot+[fontfiglisas,color=black,dashdotted,line width=2pt,mark=none,forget plot]  
   	table [x=SNR, y=ber_sims3, col sep=space] {\mySIMSmtres};	 		
   	
   	\addplot+[fontfiglisas,color=black,dashdotted,line width=2pt,mark=none,forget plot]  
   	table [x=SNR, y=ber_sims4, col sep=space] {\mySIMSmtres};	 		
   	
   	\addplot+[fontfiglisas,color=black,dashdotted,line width=2pt,mark=none,forget plot]  
   	table [x=SNR, y=ber_sims5, col sep=space] {\mySIMSmtres};	 		
   	
   	\addplot+[fontfiglisas,color=black,dashdotted,line width=2pt,mark=none,forget plot]  
   	table [x=SNR, y=ber_sims6, col sep=space] {\mySIMSmtres};	 		
   	
   	\addplot+[fontfiglisas,color=black,dashdotted,line width=2pt,mark=none,forget plot]  
   	table [x=SNR, y=ber_sims7, col sep=space] {\mySIMSmtres};	 		
 	\addplot+[only marks,line width=1pt,mark = text,color=black,scatter,forget plot,
 	scatter/use mapped color={draw opacity=1},
  	visualization depends on={value \thisrow{Name} \as \labela},
 	mark options=
 	{
 	 	text mark=\labela,
 	 	text mark as node=true,scale=2, 
 	 	text mark style={circle,fill=white, font=\bfseries,inner sep=.1pt,draw}
 	}] 
 	table[x=X,y=Y,col sep=space] {./my_figs4/nums_Nt12Nr60_it1MMSEsimples_MF_m1.dat};
	\addplot+[only marks,line width=1pt,mark = text,color=black,scatter,forget plot,
	scatter/use mapped color={draw opacity=1},
 	visualization depends on={value \thisrow{Name} \as \labela},
	mark options=
	{
		text mark=\labela,
		text mark as node=true,scale=2, 
		text mark style={circle,fill=white, font=\bfseries,inner sep=.1pt,draw}
	}] 
	table[x=X,y=Y,col sep=space] {./my_figs4/nums_Nt12Nr60_it1MMSEsimples_MF_m2.dat};
	\addplot+[only marks,line width=1pt,mark = text,color=black,scatter,forget plot,
	scatter/use mapped color={draw opacity=1},
 	visualization depends on={value \thisrow{Name} \as \labela},
	mark options=
	{
		text mark=\labela,
		text mark as node=true,scale=2, 
		text mark style={circle,fill=white, font=\bfseries,inner sep=.1pt,draw}
	}]
	table[x=X,y=Y,col sep=space] {./my_figs4/nums_Nt12Nr60_it1MMSEsimples_MF_m3.dat};
 	\end{semilogyaxis}
 	\end{tikzpicture}
	\vspace{-.25cm}\begin{quote} \hspace{0cm}\centering{\bf(a)} \end{quote}
\end{minipage}
\begin{minipage}{.5\textwidth}
		
\pgfplotstableread{./my_figs4/Nt24Nr120_it1MMSEsimples_MF_m1.dat}\mySIMSmum
\pgfplotstableread{./my_figs4/Nt24Nr120_it1MMSEsimples_MF_m2.dat}\mySIMSmdois
\pgfplotstableread{./my_figs4/Nt24Nr120_it1MMSEsimples_MF_m3.dat}\mySIMSmtres
 	
	
 	\begin{tikzpicture} 
 	
 	\begin{semilogyaxis}[
 	font=\boldmath,
 	ultra thick,
 	clip=true, 
 	xmin=-20,xmax=4,ymin=1e-4,ymax=1e-0,ymode=log,grid=both,
 	height = 7cm,
 	xtick  = {-20,-12,...,6},ytickten={-7,...,0},yticklabel pos=left,
 	xlabel = {$\tiny\bf E_b/N_0 (dB)$},	
 	ylabel = \small\bf Bit Error Rate,
 	grid   = none,
 	legend entries = {\bf Bound,\bf 4QAM,\bf 16QAM,\bf 64QAM},
 	legend style={at={(0,0)},anchor=south west,draw = none,fill=none,font=\fontsize{6}{5}\selectfont},
 	legend cell align=right,
 	]
	\tikzstyle{fontfiglisas} = [smooth,color=black,line width=1.0pt,mark options={solid,scale=1.25,fill=white}]
 	\addplot+[fontfiglisas,color=red,dotted,line width=2pt,mark=none]  
 	table [x={SNR} , y=ber_theory, col sep=space] {\mySIMSmum};
	
 	\addplot+[fontfiglisas,color=black,solid,line width=2pt,mark=none]  
 	table [x=SNR, y=ber_sims1, col sep=space] {\mySIMSmum};

 	\addplot+[fontfiglisas,color=black,solid,line width=2pt,mark=none,forget plot]  
 	table [x=SNR , y=ber_sims2, col sep=space] {\mySIMSmum};
 	
 	\addplot+[fontfiglisas,color=black,solid,line width=2pt,mark=none,forget plot]  
 	table [x=SNR, y=ber_sims3, col sep=space] {\mySIMSmum};
  	\addplot+[fontfiglisas,color=red,dotted,line width=2pt,mark=none,forget plot]  
  	table [x={SNR} , y=ber_theory, col sep=space] {\mySIMSmdois};
  	
  	\addplot+[fontfiglisas,color=black,dashed,line width=2pt,mark=none]  
  	table [x=SNR, y=ber_sims1, col sep=space] {\mySIMSmdois};
  	
  	\addplot+[fontfiglisas,color=black,dashed,line width=2pt,mark=none,forget plot]  
  	table [x=SNR , y=ber_sims2, col sep=space] {\mySIMSmdois};
  	
  	\addplot+[fontfiglisas,color=black,dashed,line width=2pt,mark=none,forget plot]  
  	table [x=SNR, y=ber_sims3, col sep=space] {\mySIMSmdois};
   	\addplot+[fontfiglisas,color=red,dotted,line width=2pt,mark=none,forget plot]  
   	table [x={SNR} , y=ber_theory, col sep=space] {\mySIMSmtres};
   	
   	\addplot+[fontfiglisas,color=black,dashdotted,line width=2pt,mark=none]  
   	table [x=SNR, y=ber_sims1, col sep=space] {\mySIMSmtres};
   	
   	\addplot+[fontfiglisas,color=black,dashdotted,line width=2pt,mark=none,forget plot]  
   	table [x=SNR , y=ber_sims2, col sep=space] {\mySIMSmtres};
   	
   	\addplot+[fontfiglisas,color=black,dashdotted,line width=2pt,mark=none,forget plot]  
   	table [x=SNR, y=ber_sims3, col sep=space] {\mySIMSmtres};	 		
   	
   	\addplot+[fontfiglisas,color=black,dashdotted,line width=2pt,mark=none,forget plot]  
   	table [x=SNR, y=ber_sims4, col sep=space] {\mySIMSmtres};	 		
   	
   	\addplot+[fontfiglisas,color=black,dashdotted,line width=2pt,mark=none,forget plot]  
   	table [x=SNR, y=ber_sims5, col sep=space] {\mySIMSmtres};	 		
   	
   	\addplot+[fontfiglisas,color=black,dashdotted,line width=2pt,mark=none,forget plot]  
   	table [x=SNR, y=ber_sims6, col sep=space] {\mySIMSmtres};	 		
 	\addplot+[only marks,line width=1pt,mark = text,color=black,scatter,forget plot,
 	scatter/use mapped color={draw opacity=1},
  	visualization depends on={value \thisrow{Name} \as \labela},
 	mark options=
 	{
 	 	text mark=\labela,
 	 	text mark as node=true,scale=2, 
 	 	text mark style={circle,fill=white, font=\bfseries,inner sep=.1pt,draw}
 	}]
 	table[x=X,y=Y,col sep=space] {./my_figs4/nums_Nt24Nr120_it1MMSEsimples_MF_m1.dat};
	\addplot+[only marks,line width=1pt,mark = text,color=black,scatter,forget plot,
	scatter/use mapped color={draw opacity=1},
 	visualization depends on={value \thisrow{Name} \as \labela},
	mark options=
	{
		text mark=\labela,
		text mark as node=true,scale=2, 
		text mark style={circle,fill=white, font=\bfseries,inner sep=.1pt,draw}
	}]
	table[x=X,y=Y,col sep=space] {./my_figs4/nums_Nt24Nr120_it1MMSEsimples_MF_m2.dat};
	\addplot+[only marks,line width=1pt,mark = text,color=black,scatter,forget plot,
	scatter/use mapped color={draw opacity=1},
 	visualization depends on={value \thisrow{Name} \as \labela},
	mark options=
	{
		text mark=\labela,
		text mark as node=true,scale=2, 
		text mark style={circle,fill=white, font=\bfseries,inner sep=.1pt,draw}
	}]
	table[x=X,y=Y,col sep=space] {./my_figs4/nums_Nt24Nr120_it1MMSEsimples_MF_m3.dat};
	\end{semilogyaxis}
 	\end{tikzpicture}
	\vspace{-.25cm}\begin{quote} \hspace{0cm}\centering{\bf(b)} \end{quote}
\end{minipage}
\begin{minipage}{.5\textwidth}
    	
\pgfplotstableread{./my_figs4/Nt12Nr120_it1MMSEsimples_MF_m1.dat}\mySIMSmum
\pgfplotstableread{./my_figs4/Nt12Nr120_it1MMSEsimples_MF_m2.dat}\mySIMSmdois
\pgfplotstableread{./my_figs4/Nt12Nr120_it1MMSEsimples_MF_m3.dat}\mySIMSmtres
 	
	
 	\begin{tikzpicture} 
 	
 	\begin{semilogyaxis}[
 	font=\boldmath,
 	ultra thick,
 	clip=true, 
 	xmin=-20,xmax=4,ymin=1e-4,ymax=1e-0,ymode=log,grid=both,
 	height = 7cm,
 	xtick  = {-20,-12,...,6},ytickten={-7,...,0},yticklabel pos=left,
 	xlabel = {$\tiny\bf E_b/N_0 (dB)$},	
 	ylabel = \small\bf Bit Error Rate,
 	grid   = none,
 	legend entries = {\bf Bound,\bf 4QAM,\bf 16QAM,\tiny\bf 64QAM},
  	legend style={at={(0,0)},anchor=south west,draw = none,fill=none,font=\fontsize{6}{5}\selectfont},
 	legend cell align=right,
 	]
	\tikzstyle{fontfiglisas} = [smooth,color=black,line width=1.0pt,mark options={solid,scale=1.25,fill=white}]
 	\addplot+[fontfiglisas,color=red,dotted,line width=2pt,mark=none]  
 	table [x={SNR} , y=ber_theory, col sep=space] {\mySIMSmum};
	
 	\addplot+[fontfiglisas,color=black,solid,line width=2pt,mark=none]  
 	table [x=SNR, y=ber_sims1, col sep=space] {\mySIMSmum};

 	\addplot+[fontfiglisas,color=black,solid,line width=2pt,mark=none,forget plot]  
 	table [x=SNR , y=ber_sims2, col sep=space] {\mySIMSmum};
  	\addplot+[fontfiglisas,color=red,dotted,line width=2pt,mark=none,forget plot]  
  	table [x={SNR} , y=ber_theory, col sep=space] {\mySIMSmdois};
  	
  	\addplot+[fontfiglisas,color=black,dashed,line width=2pt,mark=none]  
  	table [x=SNR, y=ber_sims1, col sep=space] {\mySIMSmdois};
  	
  	\addplot+[fontfiglisas,color=black,dashed,line width=2pt,mark=none,forget plot]  
  	table [x=SNR , y=ber_sims2, col sep=space] {\mySIMSmdois};
   	\addplot+[fontfiglisas,color=red,dotted,line width=2pt,mark=none,forget plot]  
   	table [x={SNR} , y=ber_theory, col sep=space] {\mySIMSmtres};
   	
   	\addplot+[fontfiglisas,color=black,dashdotted,line width=2pt,mark=none]  
   	table [x=SNR, y=ber_sims1, col sep=space] {\mySIMSmtres};
   	
   	\addplot+[fontfiglisas,color=black,dashdotted,line width=2pt,mark=none,forget plot]  
   	table [x=SNR , y=ber_sims2, col sep=space] {\mySIMSmtres};
   	
   	\addplot+[fontfiglisas,color=black,dashdotted,line width=2pt,mark=none,forget plot]  
   	table [x=SNR, y=ber_sims3, col sep=space] {\mySIMSmtres};	 		  	   	
 	\addplot+[only marks,line width=1pt,mark = text,color=black,scatter,forget plot,
 	scatter/use mapped color={draw opacity=1},
 	visualization depends on={value \thisrow{Name} \as \labela},
 	mark options=
 	{
 	 	text mark=\labela,
 	 	text mark as node=true,scale=2, 
 	 	text mark style={circle,fill=white, font=\bfseries,inner sep=.1pt,draw}
 	}]
 	table[x=X,y=Y,col sep=space] {./my_figs4/nums_Nt12Nr120_it1MMSEsimples_MF_m1.dat};
	\addplot+[only marks,line width=1pt,mark = text,color=black,scatter,forget plot,
	scatter/use mapped color={draw opacity=1},
 	visualization depends on={value \thisrow{Name} \as \labela},
	mark options=
	{
		text mark=\labela,
		text mark as node=true,scale=2, 
		text mark style={circle,fill=white, font=\bfseries,inner sep=.1pt,draw}
	}]
	table[x=X,y=Y,col sep=space] {./my_figs4/nums_Nt12Nr120_it1MMSEsimples_MF_m2.dat};
	\addplot+[only marks,line width=1pt,mark = text,color=black,scatter,forget plot,
	scatter/use mapped color={draw opacity=1},
 	visualization depends on={value \thisrow{Name} \as \labela},
	mark options=
	{
		text mark=\labela,
		text mark as node=true,scale=2, 
		text mark style={circle,fill=white, font=\bfseries,inner sep=.1pt,draw}
	}]
	table[x=X,y=Y,col sep=space] {./my_figs4/nums_Nt12Nr120_it1MMSEsimples_MF_m3.dat};
	\end{semilogyaxis}
 	\end{tikzpicture}
    \vspace{-.25cm}\begin{quote} \hspace{0cm}\centering{\bf(c)} \end{quote}
\end{minipage}
  	\caption{BER performances - when $N_T = 12$ and $N_R = 60$ (a), $N_T = 24$ and $N_R = 120$ (b) or $N_T = 12$ and $N_R = 120$ (c) - concerning an iterative DF detection, with a simplified MMSE-type detector for $p = 1$ and an MF detector for $p > 1$, for 4-QAM (solid lines), $16$-QAM (dashed lines) or $64$-QAM (dot-dashed lines) [SIMO/AWGN/MFB reference performances (dotted lines) are also included.]}  \label{Fig:simsSC_PIC_MMSEsimplesMF}
\end{figure}

Figures \ref{Fig:simsSC_PIC_MMSEsimplesMF} and  \ref{Fig:simsSC_PIC_MF} show that, by combining the use of the low-complexity linear detectors with an interference cancellation procedure, as depicted in Fig. \ref{OFDM_MIMO_Block_2}, a close approximation to the \SIMO/\MFB\ (and the practically identical \SIMO/AWGN/\MFB) can be achieved: after a few iterations, the performance gap, at $BER=10^{-3}$, becomes negligible, if required by using the simplified \MMSE-type detection - instead of the \MF\ detection - in the first iteration. By comparing the subfigures of Fig. \ref{Fig:simsSC_PIC_MMSEsimplesMF} and Fig. \ref{Fig:simsSC_PIC_MF} with the corresponding subfigures of Fig. \ref{Fig:simsSC_PIC_MF_ou_MMSEsimples}, we can observe that, after a few iterations, the low-complexity iterative DF detection can outperform the optimum (\MMSE) linear detection, which is not able to achieve a negligible performance gap regarding the SIMO/AWGN/MFB; in spite of an initial iteration where the BER performance - strongly constrained by the   $N_R/N_T$ ratio - is very poor, after several iterations we can practically achieve an ideal (interference-free and fading-free) \BER\ performance, essentially determined by $N_R$ - regardless of $N_T$ -, for each QAM scheme, provided that $N_R/N_T\geq 5$. Moreover, even when using the (extremely simple) MF detector in all iterations, the ideal BER performance can be practically achieved, even for 64-QAM schemes, when  $N_R/N_T= 10$.

Fig. \ref{simsSC_PIC_Nt24_Nr_120_MF_QAM41664QAM} shows the simulated BER performances - when  $N_T = 24$ and $N_R = 120$  - concerning the iterative \DF\ technique with an \MF\ detector for all iterations, in a specific case where different \QAM\ symbols are simultaneously transmitted through the $24$ transmitter antennas: $64$-\QAM\ for $j=1,2,3,4$, with $\Re e \{s_n^{(j)}\}$  and  $\Im m \{s_n^{(j)}\}$  equal to $\pm 1, \pm 3 , \pm 5, \pm 7$; $16$-QAM for $j=5,6,\dots,12$, with $\Re e \{s_n^{(j)}\}$  and  $\Im m \{s_n^{(j)}\}$ equal to $\pm 1, \pm 3$; $4$-QAM for $j=13,14,\dots,24$ , with  $\Re e \{s_n^{(j)}\}$  and  $\Im m \{s_n^{(j)}\}$ equal to $\pm 1$. Clearly, the \SIMO/AWGN/\MFB\ performance concerning the selected $N_R$ can be practically achieved for all \QAM\ schemes - in this realistic context where different \QAM\ schemes coexist - in spite of adopting the extremely simple detector for all iterations.

\section{Conclusions}\label{sec5}

This paper was dedicated to uplink detection issues concerning a MU-MIMO system where \QAM-based (up to 6 bits/symbol) \SC/\FDE\ transmission schemes are adopted and a large number of BS antennas is available, possibly much larger than the corresponding number of transmitter antennas at mobile terminals. In this context, we considered several detection techniques and evaluated, in detail, the 
resulting performances - discussed with the help of selected performance bounds -, for a range of values regarding the number of BS receiver antennas.

From our performance results, our main conclusion is that simple linear detection techniques, designed to avoid the need of complex matrix inversions, can lead to unacceptably high error floor levels; however, by combining the use of such simple linear detectors with an interference cancellation procedure - within a low-complexity iterative \DF\ technique -, a close approximation to the \SIMO\ \MFB\ performance can be achieved, even for $64$-\QAM\ schemes, after a few iterations, when the number of \BS\ antennas is, at least, five times higher than the number of antennas which are jointly used at the user terminals.

      \begin{figure}
      	\begin{minipage}{.5\textwidth}
      			
\pgfplotstableread{./my_figs5/Nt12Nr60_MF_m1.dat}\mySIMSmum
\pgfplotstableread{./my_figs5/Nt12Nr60_MF_m2.dat}\mySIMSmdois
\pgfplotstableread{./my_figs5/Nt12Nr60_MF_m3.dat}\mySIMSmtres
 	
 	
 	\begin{tikzpicture} 
 	
 	\begin{semilogyaxis}[
 	font=\boldmath,
 	ultra thick,
 	clip=true, 
 	xmin=-20,xmax=4,ymin=1e-4,ymax=1e-0,ymode=log,grid=both,
 	height = 7cm,
 	xtick  = {-20,-12,...,6},ytickten={-7,...,0},yticklabel pos=left,
 	xlabel = {$\tiny\bf E_b/N_0 (dB)$},	
 	ylabel = \small\bf Bit Error Rate,
 	grid   = none,
 	legend entries = {\tiny\bf 4QAM,\tiny\bf 16QAM,\tiny\bf 64QAM,\tiny\bf Bound},
 	legend style={at={(0,0)},anchor=south west,draw = none,fill=none},
 	legend cell align=right,
 	]
	\tikzstyle{fontfiglisas} = [smooth,color=black,line width=1.0pt,mark options={solid,scale=1.25,fill=white}]

 	\addplot+[fontfiglisas,color=black,solid,line width=2pt,mark=none]  
 	table [x=SNR, y=ber_sims1, col sep=space] {\mySIMSmum};

 	\addplot+[fontfiglisas,color=black,solid,line width=2pt,mark=none,forget plot]  
 	table [x=SNR , y=ber_sims2, col sep=space] {\mySIMSmum};
 	
 	\addplot+[fontfiglisas,color=black,solid,line width=2pt,mark=none,forget plot]  
 	table [x=SNR, y=ber_sims3, col sep=space] {\mySIMSmum};

 	\addplot+[fontfiglisas,color=red,dotted,line width=2pt,mark=none,forget plot]  
 	table [x={SNR} , y=ber_theory, col sep=space] {\mySIMSmum};	

  	\addplot+[fontfiglisas,color=black,dashed,line width=2pt,mark=none]  
  	table [x=SNR, y=ber_sims1, col sep=space] {\mySIMSmdois};
  	
  	\addplot+[fontfiglisas,color=black,dashed,line width=2pt,mark=none,forget plot]  
  	table [x=SNR , y=ber_sims2, col sep=space] {\mySIMSmdois};
  	
  	\addplot+[fontfiglisas,color=black,dashed,line width=2pt,mark=none,forget plot]  
  	table [x=SNR, y=ber_sims3, col sep=space] {\mySIMSmdois};
  	
  	\addplot+[fontfiglisas,color=black,dashed,line width=2pt,mark=none,forget plot]  
  	table [x=SNR, y=ber_sims4, col sep=space] {\mySIMSmdois};
  	
  	\addplot+[fontfiglisas,color=black,dashed,line width=2pt,mark=none,forget plot]  
  	table [x=SNR, y=ber_sims5, col sep=space] {\mySIMSmdois};

  	\addplot+[fontfiglisas,color=red,dotted,line width=2pt,mark=none,forget plot]  
  	table [x={SNR} , y=ber_theory, col sep=space] {\mySIMSmdois}; 	
 	  	
   	\addplot+[fontfiglisas,color=black,dashdotted,line width=2pt,mark=none]  
   	table [x=SNR, y=ber_sims1, col sep=space] {\mySIMSmtres};
   	
   	\addplot+[fontfiglisas,color=black,dashdotted,line width=2pt,mark=none,forget plot]  
   	table [x=SNR , y=ber_sims2, col sep=space] {\mySIMSmtres};
   	
   	\addplot+[fontfiglisas,color=black,dashdotted,line width=2pt,mark=none,forget plot]  
   	table [x=SNR, y=ber_sims3, col sep=space] {\mySIMSmtres};	 		
   	
   	\addplot+[fontfiglisas,color=black,dashdotted,line width=2pt,mark=none,forget plot]  
   	table [x=SNR, y=ber_sims4, col sep=space] {\mySIMSmtres};	 		
   	
   	\addplot+[fontfiglisas,color=black,dashdotted,line width=2pt,mark=none,forget plot]  
   	table [x=SNR, y=ber_sims5, col sep=space] {\mySIMSmtres};	 		
   	
   	\addplot+[fontfiglisas,color=black,dashdotted,line width=2pt,mark=none,forget plot]  
   	table [x=SNR, y=ber_sims6, col sep=space] {\mySIMSmtres};	 		
   	
   	\addplot+[fontfiglisas,color=red,dotted,line width=2pt,mark=none]  
   	table [x={SNR} , y=ber_theory, col sep=space] {\mySIMSmtres};
 	\addplot[only marks,line width=1pt,mark = text,color=black,scatter,forget plot,
 	scatter/use mapped color={draw opacity=1},
 	visualization depends on={value \thisrow{Name} \as \labela},
 	mark options=
 	{
 	 	text mark=\labela,
 	 	text mark as node=true,scale=2, 
 	 	text mark style={circle,fill=white, font=\bfseries,inner sep=.1pt,draw}
 	}]
 	table[x=X,y=Y,col sep=space] {./my_figs5/nums_Nt12Nr60_MF_m1.dat};
	\addplot[only marks,line width=1pt,mark = text,color=black,scatter,forget plot,
	scatter/use mapped color={draw opacity=1},
 	visualization depends on={value \thisrow{Name} \as \labela},
	mark options=
	{
		text mark=\labela,
		text mark as node=true,scale=2, 
		text mark style={circle,fill=white, font=\bfseries,inner sep=.1pt,draw}
	}]
	table[x=X,y=Y,col sep=space] {./my_figs5/nums_Nt12Nr60_MF_m2.dat};
	\addplot[only marks,line width=1pt,mark = text,color=black,scatter,forget plot,
	scatter/use mapped color={draw opacity=1},
 	visualization depends on={value \thisrow{Name} \as \labela},
	mark options=
	{
		text mark=\labela,
		text mark as node=true,scale=2, 
		text mark style={circle,fill=white, font=\bfseries,inner sep=.1pt,draw}
	}]
	table[x=X,y=Y,col sep=space] {./my_figs5/nums_Nt12Nr60_MF_m3.dat};
 	 	
 	 	\end{semilogyaxis}
 	\end{tikzpicture}
      		\vspace{-.25cm}\begin{quote} \hspace{.8cm}\centering{\bf(a)} \end{quote}
      	\end{minipage}
      	\begin{minipage}{.5\textwidth}
      			
\pgfplotstableread{./my_figs5/Nt24Nr120_MF_m1.dat}\mySIMSmum
\pgfplotstableread{./my_figs5/Nt24Nr120_MF_m2.dat}\mySIMSmdois
\pgfplotstableread{./my_figs5/Nt24Nr120_MF_m3.dat}\mySIMSmtres
 	
 	
 	\begin{tikzpicture} 
 	
 	\begin{semilogyaxis}[
 	font=\boldmath,
 	ultra thick,
 	clip=true, 
 	xmin=-20,xmax=4,ymin=1e-4,ymax=1e-0,ymode=log,grid=both,
 	height = 7cm,
 	xtick  = {-20,-12,...,6},ytickten={-7,...,0},yticklabel pos=left,
 	xlabel = {$\tiny\bf E_b/N_0 (dB)$},	
 	ylabel = \small\bf Bit Error Rate,
 	grid   = none,
 	legend entries = {\tiny\bf 4QAM,\tiny\bf 16QAM,\tiny\bf 64QAM,\tiny\bf Bound},
 	legend style={at={(0,0)},anchor=south west,draw = none,fill=none},
 	legend cell align=right,
 	]
	\tikzstyle{fontfiglisas} = [smooth,color=black,line width=1.0pt,mark options={solid,scale=1.25,fill=white}]

 	\addplot+[fontfiglisas,color=black,solid,line width=2pt,mark=none]  
 	table [x=SNR_Delta, y=ber_sims1, col sep=space] {\mySIMSmum};

 	\addplot+[fontfiglisas,color=black,solid,line width=2pt,mark=none,forget plot]  
 	table [x=SNR_Delta , y=ber_sims2, col sep=space] {\mySIMSmum};
 	
 	\addplot+[fontfiglisas,color=black,solid,line width=2pt,mark=none,forget plot]  
 	table [x=SNR_Delta, y=ber_sims3, col sep=space] {\mySIMSmum};
 
  	\addplot+[fontfiglisas,color=red,dotted,line width=2pt,mark=none,forget plot]  
  	table [x={SNR} , y=ber_theory, col sep=space] {\mySIMSmum};	

  	\addplot+[fontfiglisas,color=black,dashed,line width=2pt,mark=none]  
  	table [x=SNR, y=ber_sims1, col sep=space] {\mySIMSmdois};
  	
  	\addplot+[fontfiglisas,color=black,dashed,line width=2pt,mark=none,forget plot]  
  	table [x=SNR , y=ber_sims2, col sep=space] {\mySIMSmdois};
  	
  	\addplot+[fontfiglisas,color=black,dashed,line width=2pt,mark=none,forget plot]  
  	table [x=SNR, y=ber_sims3, col sep=space] {\mySIMSmdois};
  	
  	\addplot+[fontfiglisas,color=black,dashed,line width=2pt,mark=none,forget plot]  
  	table [x=SNR, y=ber_sims4, col sep=space] {\mySIMSmdois};
  	
  	\addplot+[fontfiglisas,color=black,dashed,line width=2pt,mark=none,forget plot]  
  	table [x=SNR, y=ber_sims5, col sep=space] {\mySIMSmdois};

  	\addplot+[fontfiglisas,color=red,dotted,line width=2pt,mark=none,forget plot]  
  	table [x={SNR} , y=ber_theory, col sep=space] {\mySIMSmdois}; 	
 	  	
   	\addplot+[fontfiglisas,color=black,dashdotted,line width=2pt,mark=none]  
   	table [x=SNR, y=ber_sims1, col sep=space] {\mySIMSmtres};
   	
   	\addplot+[fontfiglisas,color=black,dashdotted,line width=2pt,mark=none,forget plot]  
   	table [x=SNR , y=ber_sims2, col sep=space] {\mySIMSmtres};
   	
   	\addplot+[fontfiglisas,color=black,dashdotted,line width=2pt,mark=none,forget plot]  
   	table [x=SNR, y=ber_sims3, col sep=space] {\mySIMSmtres};	 		
   	
   	\addplot+[fontfiglisas,color=black,dashdotted,line width=2pt,mark=none,forget plot]  
   	table [x=SNR, y=ber_sims4, col sep=space] {\mySIMSmtres};	 		
   	
   	\addplot+[fontfiglisas,color=black,dashdotted,line width=2pt,mark=none,forget plot]  
   	table [x=SNR, y=ber_sims5, col sep=space] {\mySIMSmtres};	 		
   	
   	\addplot+[fontfiglisas,color=black,dashdotted,line width=2pt,mark=none,forget plot]  
   	table [x=SNR, y=ber_sims6, col sep=space] {\mySIMSmtres};	 		
   	
   	\addplot+[fontfiglisas,color=red,dotted,line width=2pt,mark=none]  
   	table [x={SNR} , y=ber_theory, col sep=space] {\mySIMSmtres};
 	\addplot[only marks,line width=1pt,mark = text,color=black,scatter,forget plot,
 	scatter/use mapped color={draw opacity=1},
 	visualization depends on={value \thisrow{Name} \as \labela},
 	mark options=
 	{
 	 	text mark=\labela,
 	 	text mark as node=true,scale=2, 
 	 	text mark style={circle,fill=white, font=\bfseries,inner sep=.1pt,draw}
 	}] 
 	table[x=X,y=Y,col sep=space] {./my_figs5/nums_Nt24Nr120_MF_m1.dat};
	\addplot[only marks,line width=1pt,mark = text,color=black,scatter,forget plot,
	scatter/use mapped color={draw opacity=1},
 	visualization depends on={value \thisrow{Name} \as \labela},
	mark options=
	{
		text mark=\labela,
		text mark as node=true,scale=2, 
		text mark style={circle,fill=white, font=\bfseries,inner sep=.1pt,draw}
	}]
	table[x=X,y=Y,col sep=space] {./my_figs5/nums_Nt24Nr120_MF_m2.dat};
	\addplot[only marks,line width=1pt,mark = text,color=black,scatter,forget plot,
	scatter/use mapped color={draw opacity=1},
 	visualization depends on={value \thisrow{Name} \as \labela},
	mark options=
	{
		text mark=\labela,
		text mark as node=true,scale=2, 
		text mark style={circle,fill=white, font=\bfseries,inner sep=.1pt,draw}
	}] 
	table[x=X,y=Y,col sep=space] {./my_figs5/nums_Nt24Nr120_MF_m3.dat};
 	 	
 	 	\end{semilogyaxis}
 	\end{tikzpicture}
      		\vspace{-.25cm}\begin{quote} \hspace{.8cm}\centering{\bf(b)} \end{quote}
      	\end{minipage}
      	\begin{minipage}{.5\textwidth}
      			
\pgfplotstableread{./my_figs5/Nt12Nr120_MF_m1.dat}\mySIMSmum
\pgfplotstableread{./my_figs5/Nt12Nr120_MF_m2.dat}\mySIMSmdois
\pgfplotstableread{./my_figs5/Nt12Nr120_MF_m3.dat}\mySIMSmtres
 	
 	
 	\begin{tikzpicture} 
 	
 	\begin{semilogyaxis}[
 	font=\boldmath,
 	ultra thick,
 	clip=true, 
 	xmin=-20,xmax=4,ymin=1e-4,ymax=1e-0,ymode=log,grid=both,
 	height = 7cm,
 	xtick  = {-20,-12,...,6},ytickten={-7,...,0},yticklabel pos=left,
 	xlabel = {$\tiny\bf E_b/N_0 (dB)$},	
 	ylabel = \small\bf Bit Error Rate,
 	grid   = none,
 	legend entries = {\tiny\bf 4QAM,\tiny\bf 16QAM,\tiny\bf 64QAM,\tiny\bf Bound},
 	legend style={at={(0,0)},anchor=south west,draw = none,fill=none},
 	legend cell align=right,
 	]
	\tikzstyle{fontfiglisas} = [smooth,color=black,line width=1.0pt,mark options={solid,scale=1.25,fill=white}]
	\addplot+[fontfiglisas,color=black,solid,line width=2pt,mark=none]  
	table [x=SNR, y=ber_sims1, col sep=space] {\mySIMSmum};

 	\addplot+[fontfiglisas,color=black,solid,line width=2pt,mark=none,forget plot]  
 	table [x=SNR , y=ber_sims2, col sep=space] {\mySIMSmum};
 	
 	\addplot+[fontfiglisas,color=black,solid,line width=2pt,mark=none,forget plot]  
 	table [x=SNR, y=ber_sims3, col sep=space] {\mySIMSmum};

 	\addplot+[fontfiglisas,color=red,dotted,line width=2pt,mark=none,forget plot]  
 	table [x={SNR} , y=ber_theory, col sep=space] {\mySIMSmum};
  	\addplot+[fontfiglisas,color=black,dashed,line width=2pt,mark=none]  
  	table [x=SNR, y=ber_sims1, col sep=space] {\mySIMSmdois};
  	
  	\addplot+[fontfiglisas,color=black,dashed,line width=2pt,mark=none,forget plot]  
  	table [x=SNR , y=ber_sims2, col sep=space] {\mySIMSmdois};
  	
  	\addplot+[fontfiglisas,color=black,dashed,line width=2pt,mark=none,forget plot]  
  	table [x=SNR, y=ber_sims3, col sep=space] {\mySIMSmdois};
  	
  	\addplot+[fontfiglisas,color=black,dashed,line width=2pt,mark=none,forget plot]  
  	table [x=SNR, y=ber_sims4, col sep=space] {\mySIMSmdois};
  	
  	\addplot+[fontfiglisas,color=black,dashed,line width=2pt,mark=none,forget plot]  
  	table [x=SNR, y=ber_sims5, col sep=space] {\mySIMSmdois};

  	\addplot+[fontfiglisas,color=red,dotted,line width=2pt,mark=none,forget plot]  
  	table [x={SNR} , y=ber_theory, col sep=space] {\mySIMSmdois};
   	\addplot+[fontfiglisas,color=black,dashdotted,line width=2pt,mark=none]  
   	table [x=SNR, y=ber_sims1, col sep=space] {\mySIMSmtres};
   	
   	\addplot+[fontfiglisas,color=black,dashdotted,line width=2pt,mark=none,forget plot]  
   	table [x=SNR , y=ber_sims2, col sep=space] {\mySIMSmtres};
   	
   	\addplot+[fontfiglisas,color=black,dashdotted,line width=2pt,mark=none,forget plot]  
   	table [x=SNR, y=ber_sims3, col sep=space] {\mySIMSmtres};	 		
   	
   	\addplot+[fontfiglisas,color=black,dashdotted,line width=2pt,mark=none,forget plot]  
   	table [x=SNR, y=ber_sims4, col sep=space] {\mySIMSmtres};	 		
   	
   	\addplot+[fontfiglisas,color=black,dashdotted,line width=2pt,mark=none,forget plot]  
   	table [x=SNR, y=ber_sims5, col sep=space] {\mySIMSmtres};	 		
   	
   	\addplot+[fontfiglisas,color=black,dashdotted,line width=2pt,mark=none,forget plot]  
   	table [x=SNR, y=ber_sims6, col sep=space] {\mySIMSmtres};	 		
   	
   	\addplot+[fontfiglisas,color=red,dotted,line width=2pt,mark=none]  
   	table [x={SNR} , y=ber_theory, col sep=space] {\mySIMSmtres};
 	\addplot[only marks,line width=1pt,mark = text,color=black,scatter,forget plot,
 	scatter/use mapped color={draw opacity=1},
 	visualization depends on={value \thisrow{Name} \as \labela},
 	mark options=
 	{
	 	 text mark=\labela,
 	 	text mark as node=true,scale=2, 
 	 	text mark style={circle,fill=white, font=\bfseries,inner sep=.1pt,draw}
 	}]
 	table[x=X,y=Y,col sep=space] {./my_figs5/nums_Nt12Nr120_MF_m1.dat};
	\addplot[only marks,line width=1pt,mark = text,color=black,scatter,forget plot,
	scatter/use mapped color={draw opacity=1},
 	visualization depends on={value \thisrow{Name} \as \labela},
	mark options=
	{
		text mark=\labela,
		text mark as node=true,scale=2, 
		text mark style={circle,fill=white, font=\bfseries,inner sep=.1pt,draw}
	}]
	table[x=X,y=Y,col sep=space] {./my_figs5/nums_Nt12Nr120_MF_m2.dat};
	\addplot[only marks,line width=1pt,mark = text,color=black,scatter,forget plot,
	scatter/use mapped color={draw opacity=1},
 	visualization depends on={value \thisrow{Name} \as \labela},
	mark options=
	{
		text mark=\labela,
		text mark as node=true,scale=2, 
		text mark style={circle,fill=white, font=\bfseries,inner sep=.1pt,draw}
	}] 
	table[x=X,y=Y,col sep=space] {./my_figs5/nums_Nt12Nr120_MF_m3.dat};
 	 	
 	 	\end{semilogyaxis}
 	\end{tikzpicture}
      		\vspace{-.25cm}\begin{quote} \hspace{.8cm}\centering{\bf(c)} \end{quote}
      	\end{minipage}
  	\caption{Similar to Fig. \ref{Fig:simsSC_PIC_MMSEsimplesMF}, but assuming an MF detector for all iterations.}  \label{Fig:simsSC_PIC_MF}
      \end{figure}
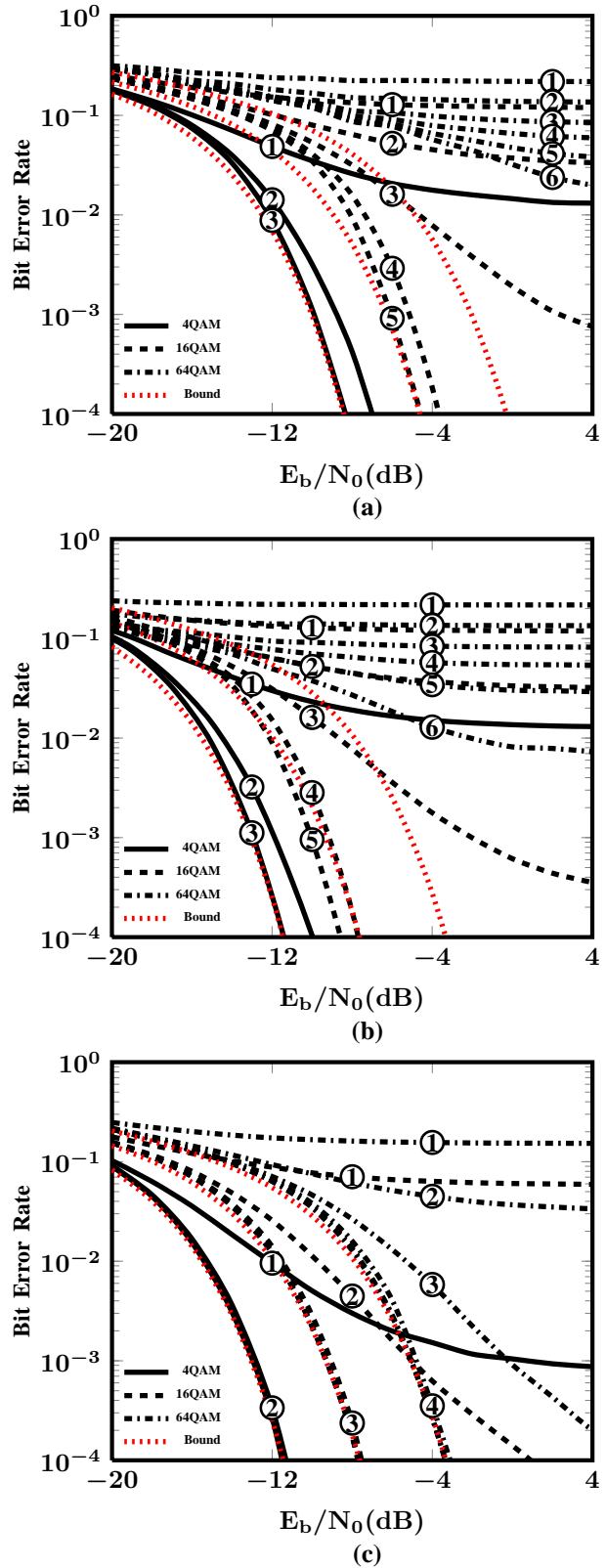
      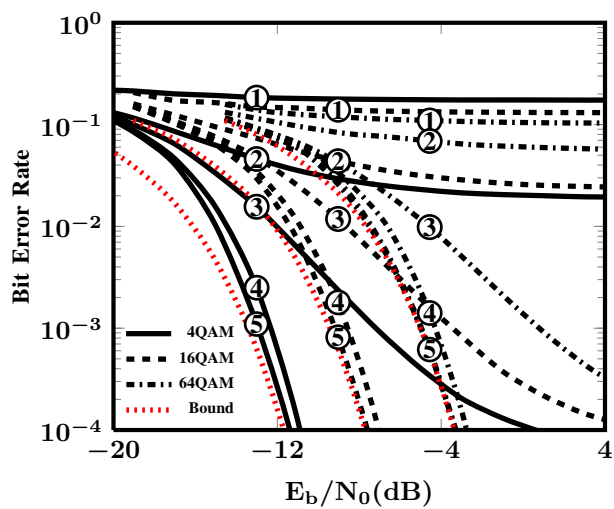
\begin{figure}
      	\begin{minipage}{.5\textwidth}
      			
\pgfplotstableread{./sims_QAM41664/Nt24Nr120_MF_QAM41664QAM_m1.dat}\mySIMSmum
\pgfplotstableread{./sims_QAM41664/Nt24Nr120_MF_QAM41664QAM_m2.dat}\mySIMSmdois
\pgfplotstableread{./sims_QAM41664/Nt24Nr120_MF_QAM41664QAM_m3.dat}\mySIMSmtres
 	
 	
 	\begin{tikzpicture} 
 	
 	\begin{semilogyaxis}[
 	font=\boldmath,
 	ultra thick,
 	clip=true, 
 	xmin=-20,xmax=4,ymin=1e-4,ymax=1e-0,ymode=log,grid=both,
 	height = 7cm,
 	xtick  = {-20,-12,...,6},ytickten={-7,...,0},yticklabel pos=left,
 	xlabel = {$\bf E_b/N_0 (dB)$},	
 	ylabel = \small\bf Bit Error Rate,
 	grid   = none,
 	legend entries = {\bf 4QAM,\bf 16QAM,\bf 64QAM,\bf Bound},
 	legend style={at={(0,0)},anchor=south west,draw = none,fill=none,font=\fontsize{6}{5}\selectfont},
 	legend cell align=right,
 	]
	\tikzstyle{fontfiglisas} = [smooth,color=black,line width=1.0pt,mark options={solid,scale=1.25,fill=white}]
 	\addplot+[fontfiglisas,color=black,solid,line width=2pt,mark=none]  
 	table [x=SNR_Delta, y=ber_sims1, col sep=space] {\mySIMSmum};

 	\addplot+[fontfiglisas,color=black,solid,line width=2pt,mark=none,forget plot]  
 	table [x=SNR_Delta , y=ber_sims2, col sep=space] {\mySIMSmum};
 	
 	\addplot+[fontfiglisas,color=black,solid,line width=2pt,mark=none,forget plot]  
 	table [x=SNR_Delta, y=ber_sims3, col sep=space] {\mySIMSmum};

 	\addplot+[fontfiglisas,color=black,solid,line width=2pt,mark=none,forget plot]  
 	table [x=SNR_Delta , y=ber_sims4, col sep=space] {\mySIMSmum};
 	
 	\addplot+[fontfiglisas,color=black,solid,line width=2pt,mark=none,forget plot]  
 	table [x=SNR_Delta, y=ber_sims5, col sep=space] {\mySIMSmum};

 	\addplot+[fontfiglisas,color=red,dotted,line width=2pt,mark=none,forget plot]  
 	table [x={SNR_Delta} , y=ber_theory, col sep=space] {\mySIMSmum};		
  	\addplot+[fontfiglisas,color=black,dashed,line width=2pt,mark=none]  
  	table [x=SNR_Delta, y=ber_sims1, col sep=space] {\mySIMSmdois};
  	
  	\addplot+[fontfiglisas,color=black,dashed,line width=2pt,mark=none,forget plot]  
  	table [x=SNR_Delta , y=ber_sims2, col sep=space] {\mySIMSmdois};
  	
  	\addplot+[fontfiglisas,color=black,dashed,line width=2pt,mark=none,forget plot]  
  	table [x=SNR_Delta, y=ber_sims3, col sep=space] {\mySIMSmdois};
  	
  	\addplot+[fontfiglisas,color=black,dashed,line width=2pt,mark=none,forget plot]  
  	table [x=SNR_Delta, y=ber_sims4, col sep=space] {\mySIMSmdois};
  	
  	\addplot+[fontfiglisas,color=black,dashed,line width=2pt,mark=none,forget plot]  
  	table [x=SNR_Delta, y=ber_sims5, col sep=space] {\mySIMSmdois};

  	\addplot+[fontfiglisas,color=red,dotted,line width=2pt,mark=none,forget plot]  
  	table [x={SNR_Delta} , y=ber_theory, col sep=space] {\mySIMSmdois}; 		
   	\addplot+[fontfiglisas,color=black,dashdotted,line width=2pt,mark=none]  
   	table [x=SNR_Delta, y=ber_sims1, col sep=space] {\mySIMSmtres};
   	
   	\addplot+[fontfiglisas,color=black,dashdotted,line width=2pt,mark=none,forget plot]  
   	table [x=SNR_Delta , y=ber_sims2, col sep=space] {\mySIMSmtres};
   	
   	\addplot+[fontfiglisas,color=black,dashdotted,line width=2pt,mark=none,forget plot]  
   	table [x=SNR_Delta, y=ber_sims3, col sep=space] {\mySIMSmtres};	 		
   	
   	\addplot+[fontfiglisas,color=black,dashdotted,line width=2pt,mark=none,forget plot]  
   	table [x=SNR_Delta, y=ber_sims4, col sep=space] {\mySIMSmtres};	 		
   	
   	\addplot+[fontfiglisas,color=black,dashdotted,line width=2pt,mark=none,forget plot]  
   	table [x=SNR_Delta, y=ber_sims5, col sep=space] {\mySIMSmtres};	 		
   	
   	\addplot+[fontfiglisas,color=black,dashdotted,line width=2pt,mark=none,forget plot]  
   	table [x=SNR_Delta, y=ber_sims6, col sep=space] {\mySIMSmtres};	 		
   	
   	\addplot+[fontfiglisas,color=red,dotted,line width=2pt,mark=none]  
   	table [x={SNR_Delta} , y=ber_theory, col sep=space] {\mySIMSmtres};   	
 	\addplot[only marks,line width=1pt,mark = text,color=black,scatter,forget plot,
 	scatter/use mapped color={draw opacity=1},
 	visualization depends on={value \thisrow{Name} \as \labela},
 	mark options=
 	{
 	 	text mark=\labela,
 	 	text mark as node=true,scale=2, 
 	 	text mark style={circle,fill=white, font=\bfseries,inner sep=.1pt,draw}
 	}]
 	table[x=X,y=Y,col sep=space] {./sims_QAM41664/nums_Nt24Nr120_MF_QAM41664QAM_m1.dat};
	\addplot[only marks,line width=1pt,mark = text,color=black,scatter,forget plot,
	scatter/use mapped color={draw opacity=1},
 	visualization depends on={value \thisrow{Name} \as \labela},
	mark options=
	{
		text mark=\labela,
		text mark as node=true,scale=2, 
		text mark style={circle,fill=white, font=\bfseries,inner sep=.1pt,draw}
	}]
	table[x=X,y=Y,col sep=space] {./sims_QAM41664/nums_Nt24Nr120_MF_QAM41664QAM_m2.dat};
	\addplot[only marks,line width=1pt,mark = text,color=black,scatter,forget plot,
	scatter/use mapped color={draw opacity=1},
 	visualization depends on={value \thisrow{Name} \as \labela},
	mark options=
	{
		text mark=\labela,
		text mark as node=true,scale=2, 
		text mark style={circle,fill=white, font=\bfseries,inner sep=.1pt,draw}
	}] 
	table[x=X,y=Y,col sep=space] {./sims_QAM41664/nums_Nt24Nr120_MF_QAM41664QAM_m3.dat};
 	 	
 	 	\end{semilogyaxis}
 	\end{tikzpicture}
      	\end{minipage}
  	\caption{BER performances - when $N_T=24$ and $N_R=120$ - concerning an iterative DF detection with an MF detector for all iterations, for $4$-QAM (solid lines), $16$-QAM (dashed lines) or $64$-QAM (dot-dashed lines), respectively, in $12$, $8$ and $4$ TX antennas  [SIMO/AWGN/MFB reference performances (dotted lines) are also included.].}   \label{simsSC_PIC_Nt24_Nr_120_MF_QAM41664QAM}
      \end{figure}
      %


\end{document}